\pdfoutput=1
\documentclass[11pt,a4paper]{article}

\usepackage{jheppub}

\usepackage{amsmath,mathtools,bm,amsfonts}

\hypersetup{
	bookmarksnumbered=true,
	bookmarksopen=true,
	bookmarksopenlevel=2
}

\usepackage{xcolor}
\usepackage{subfig}
\usepackage{float}
\usepackage{array}
\usepackage{enumitem}
\usepackage{relsize}

\numberwithin{equation}{section}

\usepackage[defaultlines=2,all]{nowidow}
\makeatletter
\def\@fpheader{\phantom{Prepared for submission to JHEP}}
\makeatother

\newcommand{\bea}{{\begin{eqnarray}}}
\newcommand{\eea}{{\end{eqnarray}}}
\newcommand{\mA}{{\mathcal A}}

\newcommand{\be}{\begin{equation}}
\newcommand{\ee}{\end{equation}}
\newcommand{\bpm}{\begin{pmatrix}}
\newcommand{\epm}{\end{pmatrix}}

\newcommand{\beqn}{\begin{eqnarray}}
\newcommand{\eeqn}{\end{eqnarray}}

\newcommand{\p}{\partial}

\newcommand{\ba}{\begin{aligned}}
\newcommand{\ea}{\end{aligned}}
\newcommand{\bi}{\begin{enumerate}}
\newcommand{\ei}{\end{enumerate}}

\newcommand{\eq}[1]{\begin{align}#1\end{align}}
\newcommand{\eqsp}[1]{\begin{equation}\begin{split}#1\end{split}\end{equation}}

\DeclareMathOperator{\arccosh}{arccosh}
\DeclareMathOperator{\arcsinh}{arcsinh}
\DeclareMathOperator{\arctanh}{arctanh}

\newcommand{\abs}[1]{\left|#1\right|}

\newcommand{\order}[1]{\mcal{O}(#1)}

\newcommand{\mbb}[1]{\mathbb{#1}}

\newcommand{\mcal}[1]{\mathcal{#1}}

\newcommand{\mop}[1]{\operatorname{#1}}
\newcommand{\thinmspace}[1][.5]{\mspace{#1\thinmuskip}}
\newcommand{\pdd}[1]{\partial_{#1}}
\newcommand{\pd}{\partial}

\def\til#1{\tilde{#1}}

\def\p{\partial}

\def\d{\delta}

\def\vp{\varphi}

\title{Extremal surfaces in glue-on AdS/$T\bar T$ holography}
\author{Luis Apolo$^{\,a}$,}
\author{Peng-Xiang Hao$^{\,b,\,c}$,}
\author{Wen-Xin Lai$^{\,b,\,d}$,}
\author{Wei Song$^{\,b,\,d}$}

\affiliation[\ensuremath{a}]{Beijing Institute of Mathematical Sciences and Applications, Beijing 101408, China}
\affiliation[\ensuremath{b}]{Yau Mathematical Sciences Center, Tsinghua University, Beijing 100084, China}
\affiliation[\ensuremath{c}]{Yukawa Institute for Theoretical Physics, Kyoto University, Kyoto 606-8502, Japan}
\affiliation[\ensuremath{d}]{Department of Mathematical Sciences, Tsinghua University, Beijing 100084, China}

\emailAdd{apolo@bimsa.cn}
\emailAdd{pxhao@yukawa.kyoto-u.ac.jp}
\emailAdd{laiwx19@mails.tsinghua.edu.cn}
\emailAdd{wsong2014@mail.tsinghua.edu.cn}

\abstract{$T\bar T$ deformed CFTs with positive deformation parameter have been proposed to be holographically dual to Einstein gravity in a glue-on $\mathrm{AdS}_3$ spacetime \cite{Apolo:2023vnm}. The latter is constructed from AdS$_3$ by gluing a patch of an auxiliary AdS$_3^*$ spacetime to its asymptotic boundary. In this work, we propose a glue-on version of the Ryu-Takayanagi formula, which is given by the signed area of an extremal surface. The extremal surface is anchored at the endpoints of an interval on a cutoff surface in the glue-on geometry. It consists of an RT surface lying in the AdS$_3$ part of the spacetime and its extension to the AdS$_3^*$ region. The signed area is the length of the RT surface minus the length of the segments in AdS$_3^*$. We find that the Ryu-Takayanagi formula with the signed area reproduces the entanglement entropy of a half interval for $T\bar T$-deformed CFTs on the sphere. We then study the properties of extremal surfaces on various glue-on geometries, including Poincar\'e $\mathrm{AdS}_3$, global $\mathrm{AdS}_3$, and the BTZ black hole. When anchored on multiple intervals at the boundary, the signed area of the minimal surfaces undergoes phase transitions with novel properties. In all of these examples, we find that the glue-on extremal surfaces exhibit a minimum length related to the deformation parameter of $T\bar T$-deformed CFTs.}

\usepackage{xspace}
\newcommand{\TTbar}{\texorpdfstring{\ensuremath{T\bar{T}}}{TTbar}\xspace}
\newcommand{\sArea}{\operatorname{\widetilde{Area}}}

\begin{document}
\maketitle

\section{Introduction} \label{se:introduction}

The AdS/CFT correspondence provides a nonperturbative formulation of quantum gravity in asymptotically anti-de Sitter spacetimes. Although originally formulated in string theory, many aspects of the correspondence are universal and sensitive only to its low energy (super)gravity approximation. As with many other examples of dualities in physics, the AdS/CFT correspondence can be used both ways, namely, it can be used to learn aspects of gravity from conformal field theory and vice versa.

In the early days of AdS/CFT, relevant deformations of the boundary CFT by a double-trace operator $\mathcal O^2$ were understood to induce a change in the boundary conditions of the field $\phi$ dual to $\mathcal O$ \cite{Klebanov:1999tb,Witten:2001ua}. Since this deformation induces an RG flow from a UV to an IR fixed point, the asymptotically AdS metric is not affected by the deformation. A more dramatic effect is found, however, when the boundary CFT is deformed by an irrelevant operator involving components of the stress tensor, e.g.~by the $T\bar T$ operator \cite{Zamolodchikov:2004ce,Smirnov:2016lqw,Cavaglia:2016oda}. In this case, the boundary conditions of an otherwise asymptotically AdS spacetime are changed, and mix both leading and subleading components of the metric \cite{Guica:2019nzm}. 

The $T\bar T$ deformation of a holographic CFT$_2$ can be used to gain a better understanding of holography as we move away from strictly asymptotically AdS$_3$ spacetimes.\footnote{There is a single-trace version of the $T\bar T$ deformation, which is not the subject of the present paper, and is obtained from the symmetric product orbifold of a $T\bar T$-deformed CFT. These theories are holographically related to string theory on TsT backgrounds which are no longer asymptotically AdS$_3$ \cite{Giveon:2017nie,Borsato:2018spz,Araujo:2018rho,Apolo:2019zai}.} Conversely, gravity can be used to gain a better understanding of holographic CFTs deformed by the $T\bar T$ operator. This approach is particularly appealing because it gives a geometric interpretation to the $T\bar T$ deformation. 

When the deformation parameter $\mu$ is negative, the $T\bar T$ deformation induces Dirichlet boundary conditions for the metric on a hypersurface at a fixed radial distance from the origin of the spacetime \cite{McGough:2016lol}.\footnote{Note that this version of holography is valid in the absence of bulk matter fields, i.e.~for pure gravity.} This is equivalent to introducing a finite cutoff in the bulk and the $T\bar T$-deformed CFT can be interpreted as living at this cutoff surface. The advantage of this formulation is that many results in $T\bar T$-deformed CFTs with $\mu < 0$ can be understood geometrically as a consequence of the finite cutoff. This includes, in particular, the derivation of the spectrum, superluminal propagation, black hole thermodynamics, and its partition function \cite{McGough:2016lol,Apolo:2023vnm}.

\begin{figure}[!ht]
\centering
 \includegraphics[scale=0.478]{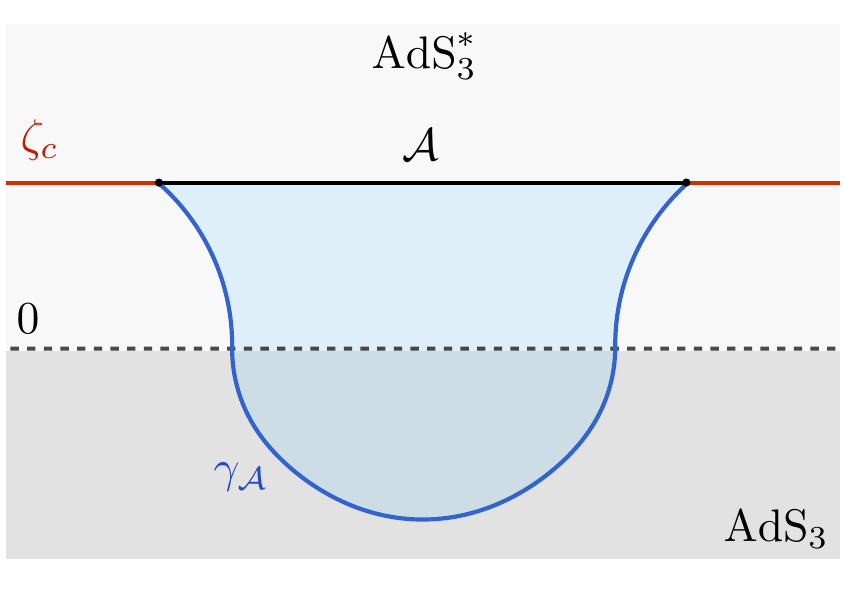}
\caption{A glue-on AdS$_3$ spacetime consists of two locally AdS$_3$ spacetimes, denoted  by AdS$_3$ and AdS$_3^*$, glued along their asymptotic boundaries at $\zeta = 0$. The glue-on HRT surface $\gamma_\mA$ (blue) associated with the interval $\mA$ on the cutoff surface (red) at $\zeta = \zeta_c$ consists of a standard HRT surface in AdS$_3$ and two hyperbolic segments in AdS$_3^*$.}
\label{fig:poincareRT_intro}
\end{figure}

The sign of the $T\bar T$ deformation plays a crucial role in the properties of the theory. For example, a negative value of $\mu$ leads to superluminal propagation and a complex spectrum at high energies. These problems are not present when $\mu > 0$, case in which $T\bar T$-deformed CFTs are well defined for arbitrarily high energies. The cutoff AdS$_3$ proposal is not applicable for $\mu > 0$, however. In this case, we have put forward a new holographic proposal dubbed \emph{glue-on} AdS$_3$ holography. In this proposal, the cutoff surface is pushed beyond the asymptotic boundary of AdS$_3$ into an auxiliary AdS$_3^*$ region. The $T\bar T$-deformed CFT$_2$ can be viewed as living on this cutoff surface \cite{Apolo:2023vnm} (see fig.~\ref{fig:poincareRT_intro}). We have previously shown that glue-on AdS$_3$ holography reproduces the spectrum, subluminal propagation, black hole thermodynamics, and the partition function of $T\bar T$-deformed CFTs with a positive deformation parameter.

In AdS/CFT, the Ryu-Takayanagi and Hubeny-Rangamani-Takayanagi (HRT) formulae tell us that the area of an extremal surface attached to the boundary of an interval at the AdS boundary yields the entanglement entropy of that interval in the dual CFT \cite{Ryu:2006bv,Hubeny:2007xt}. The HRT formula is an example of how geometry encodes features of the dual CFT and it has been instrumental in shaping our understanding of the emergence of spacetime. It is therefore natural to ask what role do HRT surfaces play in the glue-on AdS$_3$/$T\bar T$ correspondence. This question is particularly interesting because, with the exception of the $\mu < 0$ results on the sphere and on the plane \cite{Donnelly:2018bef,Lewkowycz:2019xse}, a general non-perturbative derivation of the entanglement entropy of $T\bar T$-deformed CFTs is still lacking (see \cite{Chakraborty:2018kpr,Chen:2018eqk,Park:2018snf,Murdia:2019fax,Jeong:2019ylz,Asrat:2020uib,He:2023xnb,Ashkenazi:2023fcn,Tian:2023fgf,Hou:2023ytl,Castro-Alvaredo:2023jbg} for related perturbative and nonperturbative approaches). 

In this paper we study extremal surfaces in glue-on AdS$_3$ spacetimes. Given an interval $\mA$ on the cutoff surface,  the glue-on version of the HRT formula is proposed to be given by the minimum value of the signed area of spacelike extremal surfaces homologous to $\mA$. This can be written as
\eq{
\tilde S[\mA] =
\operatorname*{\min \mop{ext}}
\limits_{\vphantom{\hat{X}}\! X_\mA\sim\mA} {\sArea[X_\mA]\over4G},
\label{proposal_intro}
}
where $G$ is Newton's constant, the spacelike surfaces $X_\mA = X \cup X^*$ lie on both the AdS$_3$ and AdS$_3^*$ regions, and $X_\mA \sim \mA$ denotes all the surfaces $X_\mA$ homologous to $\mA$. The $\sArea[X_\mA]$ is the signed area of $X_\mA$, which is the difference between the lengths of its AdS$_3$ and AdS$_3^*$ parts such that\footnote{The glue-on HRT proposal is reminiscent of the swing surface proposal for the entanglement entropy of warped CFTs and BMS-invariant quantum field theories put forward in \cite{Jiang:2017ecm,Apolo:2020bld}. In analogy with that construction, there is a spacelike surface in the bulk that is attached to two segments that connect it to the endpoints of $\mA$. In contrast to swing surfaces, where the two segments are null and do not contribute to the area, the two segments of a glue-on surface are hyperbolic, and contribute to the area via \eqref{signed_area_intro}.}
\eq{
\sArea[X_\mA] = \textrm{Area}[X] - \textrm{Area}[X^*]. \label{signed_area_intro}
}
As shown in explicit examples, spacelike surfaces that extremize the signed area do not always exist.  When there is no extremal surface for an interval $\mA$, we define the glue-on HRT formula as $\tilde S[\mA]=0$. When extremal surfaces exist, the glue-on HRT formula is given by $\tilde S[\mA]={\sArea[\gamma_\mA]\over 4G}$, where $\gamma_\mA$ is 
 the extremal surface that minimizes the signed area. The glue-on HRT surface $\gamma_\mA$ consists of two parts: a standard HRT surface $\gamma$ that lies entirely in AdS$_3$, and its extension  $\gamma^{*}$ to the AdS$_3^*$ region where it attaches to the endpoints of $\mA$. This is illustrated in fig.~\ref{fig:poincareRT_intro}.

The simplest glue-on HRT surface we consider is obtained when the interval $\mA$ connects two antipodal points on a two-sphere. In this case, the bulk spacetime is the glue-on extension of the sphere foliation of Euclidean AdS$_3$. The corresponding glue-on HRT surface consists of a straight line through the center of the space that connects two antipodal points of the sphere. In this case, the signed area of the glue-on HRT surface matches the entanglement entropy $S[\mathcal A]$ of an interval $\mA$ in a $T\bar T$-deformed CFT with a positive deformation parameter \cite{Donnelly:2018bef}\footnote{The matching to the entanglement entropy of $T\bar T$-deformed CFTs works provided that the deformation parameter is identified with the UV cutoff of the theory. Also note that a holographic derivation of the entropy in the case $\mu < 0$ was obtained from the length of an HRT surface in cutoff AdS$_3$ in \cite{Donnelly:2018bef}.}
\eq{
\tilde S[\mA] =\frac{c}{3} \arccosh \bigg(\sqrt{\frac{3}{c\mu}}\frac{\ell_\mA}{\pi} \bigg)= S[\mA],\qquad \mu>0. 
}
This motivates our study of extremal surfaces in more general cases with the ultimate goal of understanding their relationship to the entanglement entropy of $T\bar T$-deformed CFTs in more general scenarios.

In order to further understand the glue-on HRT proposal \eqref{proposal_intro}, we work out explicit examples in Poincar\'e AdS$_3$, global AdS$_3$, and BTZ black holes. We expect this novel geometrical quantity to be related to the entanglement entropy of $T\bar T$-deformed CFTs, and our results are compatible with this expectation. We can show that the glue-on HRT formula in Poincar\'e AdS$_3$ has several nice features including non-negativity, monotonicity, concavity, purity, and the infinitesimal version of strong subadditivity. We also find a minimum length $\ell^2_{\min}=\frac{4c\mu}{3}$ below which extremal surfaces do not exist, so that the signed area vanishes by definition. For example, the glue-on HRT formula for a single interval $\mA$ in  Poincar\'e AdS$_3$ reads
\eq{
\tilde S[\mA] =\frac{\sArea[\gamma_\mA]}{4G}=\left\lbrace
\begin{aligned} 
\,&0,&& \ell_{\mA} \le \ell_{\min},\\
&
\frac{c}{3} \arccosh\biggl( \sqrt{\frac{3}{\smash[b]{ c\mu }}} \frac{\ell_\mA}{2} \thinmspace  \biggr),\quad &&\ell_{\mA} >\ell_{\min}.
\end{aligned} \label{eq:HEEPoincareintro}
\right.
}
The appearance of a minimum length is compatible with our expectations from the \TTbar deformation where the deformation parameter $\mu$ is related to a minimum length, see e.g.~\cite{Dubovsky:2012wk}. The single interval result \eqref{eq:HEEPoincareintro} is a building block for multiple intervals. In this case, the signed area of the glue-on HRT surfaces undergoes a phase transition that is similar to that of a CFT$_2$ with a finite cutoff. The existence of a minimum length also plays an important role here, as it can lead to violations of subadditivity and strong subadditivity.

The paper is organized as follows. In section \ref{se:glueonreview} we review the $T\bar T$ deformation and glue-on AdS$_3$ holography. In section \ref{se:ERTsphere} we consider the entanglement entropy of $T\bar T$-deformed CFTs on a half interval on the sphere and show how this result can be reproduced from the signed area of a glue-on HRT surface. In section \ref{se:hrt} we provide a general glue-on HRT formula. Therein we describe in detail the glue-on HRT surface for a single interval in Poincar\'e AdS$_3$, discuss the emergence of a minimum length, and find the phase diagram for the signed area of surfaces associated with multiple intervals. In this section we also describe general properties of the glue-on HRT formula. In section \ref{s5} we construct glue-on HRT surfaces in global AdS$_3$ and BTZ spacetimes, and describe some of their properties. In appendix \ref{se:renormailzed-EE} we consider a more general prescription for the holographic entanglement entropy of a half interval on the sphere, and in appendix \ref{se:general-cylinder} we provide a general formula for the signed area of an extremal surface on arbitrary stationary solutions of Einstein gravity.


\section{Glue-on AdS holography for \TTbar-deformed CFTs} \label{se:glueonreview}

In this section we review the $T\bar T$ deformation and its holographic description in terms of glue-on AdS$_3$ spacetimes.

\subsection{\TTbar-deformed CFTs}

The $T\bar T$ deformation of a two-dimensional QFT is a solvable irrelevant deformation driven by the stress tensor $T_{ij}$ such that the action $I$ satisfies \cite{Zamolodchikov:2004ce,Smirnov:2016lqw,Cavaglia:2016oda}
\eq{ \label{TTbardef}
\p_\mu I = 8\pi \int d^2x \sqrt{-\gamma} \, T\bar T = \pi\int d^2x \sqrt{-\gamma} \,\big( T^{ij}T_{ij}- (T^i_i)^2 \big) ,
}
where $\mu$ is the deformation parameter and the stress tensor is defined by
\eq{
 T_{ij} = \frac{2}{\sqrt{-\gamma}} \frac{\d I}{\d \gamma^{ij}} ,
}
where $\gamma_{ij}$ is the metric of the spacetime the QFT is defined on. The $T \bar T$ deformation preserves the translational symmetry of the undeformed theory such that
\eq{
\nabla_i \langle T^i\,_j\rangle=0. \label{diffT}
}
Furthermore, when the undeformed theory is a CFT with central charge $c$, the trace of the stress tensor satisfies the trace-flow equation \cite{McGough:2016lol,Shyam:2017znq,Hartman:2018tkw,Guica:2019nzm}
\eq{
\langle T^i_i \rangle = -\frac{c}{24\pi} R^{(2)}
+ 16\pi\mu\,\langle T\bar T\rangle, \label{floweq}
}
where $R^{(2)}$ is the Ricci scalar of the background metric $\gamma_{ij}$.

The \TTbar deformation enjoys a number of properties that make it attractive from a purely field theoretical point of view, see \cite{Jiang:2019epa} for a review. In particular,  the expectation value of the \TTbar operator is finite, i.e.~free of coincident-point singularities, and it factorizes into the product of expectation values of the stress tensor \cite{Zamolodchikov:2004ce}. The factorizability of the \TTbar operator implies that the spectrum of \TTbar-deformed CFTs on the cylinder is solvable and given by
\eq{
E(\mu) = - \frac{R}{2\mu} \bigg( 1 - \sqrt{1 + \frac{4\mu}{R} E(0) + \frac{4 \mu^2}{R^2} J(0)^2}\,\bigg),
}
where $R$ is the size of the cylinder, $E(0)$ is the undeformed energy, and $J(0)$ is the undeformed angular momentum. When the deformation parameter is negative, the argument of the square root becomes negative for large enough $E(0)$. As a result, the spectrum of \TTbar-deformed CFTs with $\mu < 0$ becomes complex at high energies. In contrast, the spectrum of \TTbar-deformed CFTs with $\mu > 0$ is well defined at all energies provided that the deformation parameter is bounded by
\eq{
\mu \le \frac{3 R^2}{c}. \label{criticalmu}
}
This bound for $\mu$ arises by requiring a real ground state energy. For positive $\mu$ satisfying \eqref{criticalmu}, the torus partition function is well defined and shown to be modular invariant \cite{Datta:2018thy}. Furthermore, modular invariance implies that the torus partition function is universal when the undeformed CFT has a large central charge and a sparse spectrum~\cite{Apolo:2023aho}.


\subsection{Glue-on AdS holography} \label{se:glueonAdS}

Let us consider a two-dimensional CFT that is dual to three-dimensional Einstein gravity with a negative cosmological constant. The AdS/CFT dictionary \cite{Klebanov:1999tb,Witten:2001ua} tells us that deforming the CFT by the \TTbar operator changes the boundary conditions of the bulk metric $g_{\mu\nu}$ \cite{Guica:2019nzm}. In the absence of bulk matter fields, there is an alternative holographic description where the metric satisfies Dirichlet boundary conditions at a cutoff surface in the bulk. When $\mu < 0$, this cutoff surface is located in the interior of the spacetime \cite{McGough:2016lol}. On the other hand, when $\mu > 0$ the cutoff surface is located in an auxiliary AdS$_3^*$ region that is obtained from analytic continuation of AdS$_3$ and glued to its asymptotic boundary \cite{Apolo:2023vnm}. This version of holography for \TTbar-deformed CFTs is dubbed glue-on AdS$_3$ holography.

In order to describe the glue-on version of holography in more detail, let us consider the foliation of locally AdS$_3$ spacetimes by timelike surfaces $\mathcal N_\zeta$ with a constant radial function $\zeta(x^{\mu})$. The metric can be written in terms of the coordinates $x^\mu = (\zeta, x^i)$ as
\eq{
\text{AdS}_3: \quad ds^2 = n_\mu n_\nu dx^\mu dx^\nu + {1\over \zeta} \gamma_{ij}dx^i dx^j,\qquad \zeta>0,  \label{se2:ads3}
} 
where $n^\mu$ is the unit vector normal to $\mathcal N_\zeta$ and $x^i$ are the coordinates on $\mathcal N_\zeta$. In this gauge, the asymptotic boundary of AdS$_3$ is located at $\zeta \to 0^+$ and points with $\zeta>0$ are located in the interior of the AdS$_3$ spacetime. For the spacetime to be asymptotically AdS$_3$, the normal-normal component of the metric must have a fixed leading order falloff so that \cite{Brown:1986nw}
\eq{
n_\mu n_\nu dx^\mu dx^\nu
= \frac{d\zeta^2}{4\zeta^2} + \mathcal{O}(\zeta^{-1}),
\qquad \zeta \to 0.
\label{eq:ads-falloff}
}

The \TTbar deformation of a two-dimensional CFT is proposed to be holographically dual to Einstein gravity on a cutoff \cite{McGough:2016lol} or glue-on \cite{Apolo:2023vnm} AdS$_3$ spacetime defined by 
\eq{
\text{cutoff/glue-on AdS}_3: \quad  ds^2 = n_\mu n_\nu dx^\mu dx^\nu + {1\over \zeta} \gamma_{ij}dx^i dx^j,\qquad \zeta \ge \zeta_c , \label{se2:glueonAdS}
}
with the metric satisfying Dirichlet boundary conditions at the cutoff surface $\mathcal N_c\colon \zeta = \zeta_c$. The location of the cutoff surface $\mathcal N_c$ is related to the $T\bar T$ deformation parameter $\mu$ by
\eq{
\zeta_c \equiv - \frac{c \mu}{3\ell^2}. \label{dictionary}
}
When $\mu<0$, the cutoff surface $\mathcal{N}_c$ is moved towards the interior of AdS$_3$ and we obtain a cutoff AdS$_3$ spacetime. On the other hand, when $\mu>0$, the spacetime \eqref{se2:glueonAdS} is known as glue-on AdS$_3$ and it is obtained by analytic continuation of AdS$_3$ to negative values of $\zeta$.  

Let us denote the $\zeta<0$ region of \eqref{se2:glueonAdS} by AdS$_3^*$, which is still a locally AdS$_3$ geometry. The glue-on AdS$_3$ spacetime can be interpreted as gluing AdS$_3^*$ to the original AdS$_3$ background along each of these spacetimes' asymptotic boundaries. The crucial difference between AdS$_3$ and AdS$_3^*$ is the relative sign between the $\gamma_{ij} dx^i dx^j$ and $\zeta^{-1} \gamma_{ij} dx^i dx^j$ line elements that stems from the different signs of $\zeta$. This relative sign is telling us that the timelike coordinate $x^0$ of AdS$_3$ is spacelike in AdS$_3^*$, while the spacelike coordinate $x^1$ is timelike. Nevertheless, note that the metric the \TTbar-deformed CFT couples to is identified with $\gamma_{ij}$ such that $x^0$ is timelike and $x^1$ is spacelike for either sign of $\mu$. For more details on glue-on AdS3 holography, including evidence for the correspondence, see \cite{Apolo:2023vnm}.


\section{\TTbar holographic entanglement entropy on the sphere} \label{se:ERTsphere}

In this section we consider the holographic entanglement entropy of $T\bar T$-deformed CFTs on the sphere.  For $\mu < 0$, the entanglement entropy of an interval connecting two antipodal points of the sphere can be computed non-perturbatively from the field theory side \cite{Donnelly:2018bef}. This result can be reproduced holographically from the length of an HRT surface connecting two antipodal points of a sphere at a finite cutoff in the bulk. We will show that this result can be easily extended to the $\mu > 0$ case. On the field theory side, the entanglement entropy is found to be well defined for spheres whose radii are greater than a minimum value set by $\mu$. On the bulk side, we find that the entanglement entropy is given by the signed area of a glue-on HRT surface that connects two antipodal points on a cutoff surface in the  AdS$_3^*$ region of glue-on AdS$_3$.

\subsection{Field theory derivation} \label{se:sphere-field-theory}

In this section we calculate the entanglement entropy of a half interval on the sphere in $T\bar T$-deformed CFTs with either sign of $\mu$. The sphere partition function with $\mu < 0$ has been previously computed in \cite{Donnelly:2018bef,Li:2020zjb}, and the result has been refined and generalized for $\mu > 0$ in \cite{Apolo:2023vnm}. Here, we briefly review the derivation of the sphere partition function and then use it to calculate the entanglement entropy of a half interval.

Let us consider a $T\bar T$-deformed CFT defined on a sphere of radius $L$ with metric
\eq{
ds^2 = L^2 (d\theta^2 + \sin^2\theta \,d\phi^2).
}
The conservation law \eqref{diffT} and the trace-flow equation \eqref{floweq} can be directly solved  on the sphere, such that the stress tensor is given by~\cite{Donnelly:2018bef}
\eq{\label{tracesol}
\langle T_{ij}\rangle =-\frac{1}{4\pi\mu} \bigg(1-\sqrt{1-\frac{c\mu}{3L^2}}\, \bigg)  \gamma_{ij}.
}
Note that this expression is valid for both signs of $\mu$. In particular, for positive $\mu$, the stress tensor is real provided that the radius of the sphere is greater than a minimum value, 
\eq{
L\ge L_{\min} \equiv \sqrt{\frac{c\mu}{3}}. \label{minradius}
}

The sphere partition function depends on both the deformation parameter $\mu$ and the radius $L$. The dependence of the partition function on $\mu$ can be determined from the definition of the \TTbar deformation \eqref{TTbardef} and the flow equation \eqref{floweq},  
\eq{
\mu\thinmspace\partial_\mu \log Z_{\TTbar}(\mu)=8\pi \mu \int d^2x\sqrt{\gamma}\, \langle T\bar{T} \rangle={1\over2} \int d^2x\sqrt{\gamma}\,\Big(\langle T^i_i \rangle+{c\over 24} R^{(2)} \Big). \label{dZdmu}
}
On the other hand, a change in the radius of the sphere is equivalent to a scale transformation, the latter of which is generated by the trace of the stress tensor. As a result, the partition function satisfies the differential equation
\eq{
L\thinmspace\pdd{L} \log Z_\mu &= -\int d^2x\sqrt{\gamma}\,\langle T^i_i \rangle.
\label{dZdL}
}
Given the solution of the stress tensor \eqref{tracesol}, the general solution to the differential equations \eqref{dZdmu} and \eqref{dZdL} is then given by
\eq{\log Z_\mu(a) = \frac{c}{3} \log \bigg[\frac{L}{a}   \bigg(1+\sqrt{1-\frac{c \mu }{3  L^2}}\, \bigg) \bigg] - \frac{L^2}{\mu}  \sqrt{1-\frac{c \mu }{3 L^2}} + \frac{L^2}{\mu}, \label{Zsol}
}
where $a$ is an arbitrary integration constant with the dimension of length that is interpreted as the renormalization scale. In \cite{Apolo:2023vnm}, the cutoff scale $a$ is kept independent of the deformation parameter $\mu$, and shown to agree with the bulk on-shell action. This differs from the choice made in \cite{Donnelly:2018bef} which only depends on $\mu$. Nevertheless, the latter can be reproduced from \eqref{Zsol} by choosing the cutoff scale for $\mu<0$ to be
\eq{
\qquad a= \sqrt{\frac{c\abs{\mu}}{3}}. \label{aL}
} 
This choice is motivated by the cutoff AdS$_3$ proposal and the UV/IR relation of the AdS/CFT correspondence.\footnote{Note that in this case, the trace-flow equation \eqref{floweq} receives an additional contribution from the Weyl anomaly that originates from the rescaling of the UV cutoff induced by the flow of $\mu$ \cite{Hartman:2018tkw}.} For $\mu>0$, the choice of the cutoff scale \eqref{aL} has a natural field theoretical interpretation as the minimum length $L_{\min}$ required by the reality condition of the stress tensor \eqref{minradius}. Using \eqref{aL}, the partition function reads
\eq{
\log Z_\mu
&= \left\lbrace
\begin{aligned}
& \frac{c}{3} \arccosh \bigg(\sqrt{\frac{3}{c\mu}} L\bigg) + \frac{L^2}{\mu} \bigg(1 - \sqrt{1 - \frac{c \mu}{3L^2}} \bigg), && \mu > 0,
	\\[.5ex]
& \frac{c}{3} \arcsinh \bigg(\sqrt{-\frac{3}{c\mu}} L\bigg) + \frac{L^2}{\mu} \bigg(1 - \sqrt{1 - \frac{c \mu}{3L^2}} \bigg), \quad && \mu < 0,
\end{aligned}\label{partionsphere}
 \right.
}
which reproduces the result of \cite{Donnelly:2018bef} when $\mu <0$. 

Let us now turn to the entanglement entropy. The simplest interval $\mA$ on the sphere is the geodesic connecting two antipodal points such that its length is given by $\ell_\mA = \pi L$. The advantage of this choice is that the vacuum entanglement entropy $S[\mathcal A]$ can be easily computed using the replica trick on the $n$-sheeted sphere 
\eq{\label{nsphere}
ds^2 = L^2 (d\theta^2 + n^2 \sin^2\theta \,d\phi^2),
}
such that
\eq{
S[\mA]
= \bigg(1 - n \frac{d}{dn} \bigg) \log Z^{(n)}_\mu \big|_{n = 1},
\label{eq:replica-trick}
}
where $Z^{(n)}_\mu$ is the \TTbar partition function on \eqref{nsphere} with $Z^{(1)}_\mu = Z_\mu$. As shown in \cite{Donnelly:2018bef}, the partition function satisfies
\eq{
\partial_n\log Z_\mu \big|_{n = 1}
= -\int  dx^2 \sqrt{\gamma}\, \langle{T^\phi_\phi}\rangle
= -{1\over2}\int dx^2 \sqrt{\gamma}\, \langle T^i_i\rangle
={1\over 2}\,L\thinmspace\partial_L\log Z_\mu,
} 
where the last equality follows from \eqref{dZdL}. As a result, the entanglement entropy is determined by the partition function of $T\bar T$-deformed CFTs on the sphere via
\eq{
S[\mA] = \bigg(1 - \frac{L}{2} \frac{d}{dL} \bigg) \log Z_\mu.
}
Using the expression for the partition function in \eqref{partionsphere}, the entanglement entropy of the interval connecting two antipodal points on the sphere is thus given by
\eq{
S[\mA] = \left\lbrace
\begin{aligned}
& \frac{c}{3} \arccosh \bigg(\sqrt{\frac{3}{c\mu}} \frac{\ell_\mA}{\pi} \bigg), && \mu > 0,
	\\[.5ex]
& \frac{c}{3} \arcsinh \bigg(\sqrt{-\frac{3}{c\mu}}\frac{\ell_\mA}{\pi} \bigg), \quad && \mu < 0.
\end{aligned}
 \right. \label{EEsphere}
}
Note that in the $\mu > 0$ case, the entanglement entropy inherits the same range of validity as the partition function, being well defined only when $\ell_\mA = \pi L\ge \pi L_{\min}$. In contrast, when $\mu < 0$, we do not see the appearance of a minimum length.  In the following section we will show that the entanglement entropy for $\mu > 0$ matches a glue-on version of the HRT formula.

It is important to note that the entanglement entropy \eqref{EEsphere} does not contain an independent UV cutoff scale, which makes it impossible to reproduce the standard CFT$_2$ result in the undeformed limit $\mu\to0$. The reason is that we have already identified the deformation parameter $\mu$ with the cutoff scale, as indicated in \eqref{aL}. In appendix \ref{se:renormailzed-EE} we discuss the renormalized version of the entanglement entropy with an arbitrary choice of $a$ in the field theory side, and carry out a bulk computation to reproduce it. In the main text, we focus on the result \eqref{EEsphere}.


\subsection{Holographic entanglement entropy} \label{se:HEEsphere}

Let us now turn to the bulk side of the glue-on AdS$_3$/\TTbar correspondence. In AdS$_3$/CFT$_2$, the geometry dual to the vacuum of a CFT on the sphere is given by the sphere foliation of Euclidean AdS$_3$, which can be written as
\eq{
\text{AdS}_3: \quad ds^2 = \ell^2 \bigg[ \frac{d\zeta^2}{4\zeta^2(1+\zeta)} +  \zeta^{-1} \big( d\theta^2 + \sin\theta^2\,d\phi^2\big) \bigg], \qquad \zeta > 0. \label{adsSphere}
}
This is the same radial function used in \eqref{se2:glueonAdS} and related to the standard radial coordinate $r$ by $\zeta=1/r^2$. The glue-on version of this space is obtained by analytically continuing $\zeta$ to negative values such that
\eq{
\text{glue-on AdS}_3\colon \quad ds^2 = \ell^2 \bigg[ \frac{d\zeta^2}{4\zeta^2(1+\zeta)} +  \zeta^{-1} \big( d\theta^2 + \sin\theta^2\,d\phi^2\big) \bigg], \qquad \zeta\ge \zeta_c, \label{glueSphere}
}
where the cutoff $\zeta_c$ is related to the deformation parameter by \eqref{dictionary}, which is reproduced here for convenience
\eq{
\zeta_c = - \frac{c \mu}{3\ell^2}.
\label{dic}
}
The metric at the cutoff surface $\mathcal N_c\colon \zeta=\zeta_c$ is, by construction, a sphere of radius $L=\ell$ such that
\eq{
ds^2\big|_{\zeta_c} =\zeta_c^{-1} \, ds_c^2, \qquad ds_c^2 = \ell^2 \big( d\theta^2 + \sin\theta^2\,d\phi^2 \big). \label{boundarymetricsphere}
}

Depending on the sign of $\mu$, the cutoff surface is located either in the interior of the AdS$_3$ ($\zeta > 0$) or AdS$_3^*$ ($\zeta < 0$) parts of the spacetime. Note that the determinant of the metric in AdS$_3^*$ is positive, although both the $\theta$ and $\phi$ coordinates are now timelike. Nevertheless, the line element of the $T\bar T$-deformed theory is identified with $ds_c^2$ \eqref{boundarymetricsphere}, where both the $\theta$ and $\phi$ coordinates are spacelike. Furthermore, note that the signature of \eqref{glueSphere} changes when $\zeta < -1$, so the radial coordinate in the AdS$_3^*$ region is restricted to $\zeta \ge -1$. The holographic dictionary \eqref{dic} then implies that $\ell^2 \ge c\mu/3$, which reproduces the condition on the radius of the sphere found on the field theory side \eqref{minradius}. 

\begin{figure}[ht!]
\begin{center}
 \includegraphics[scale=0.478]{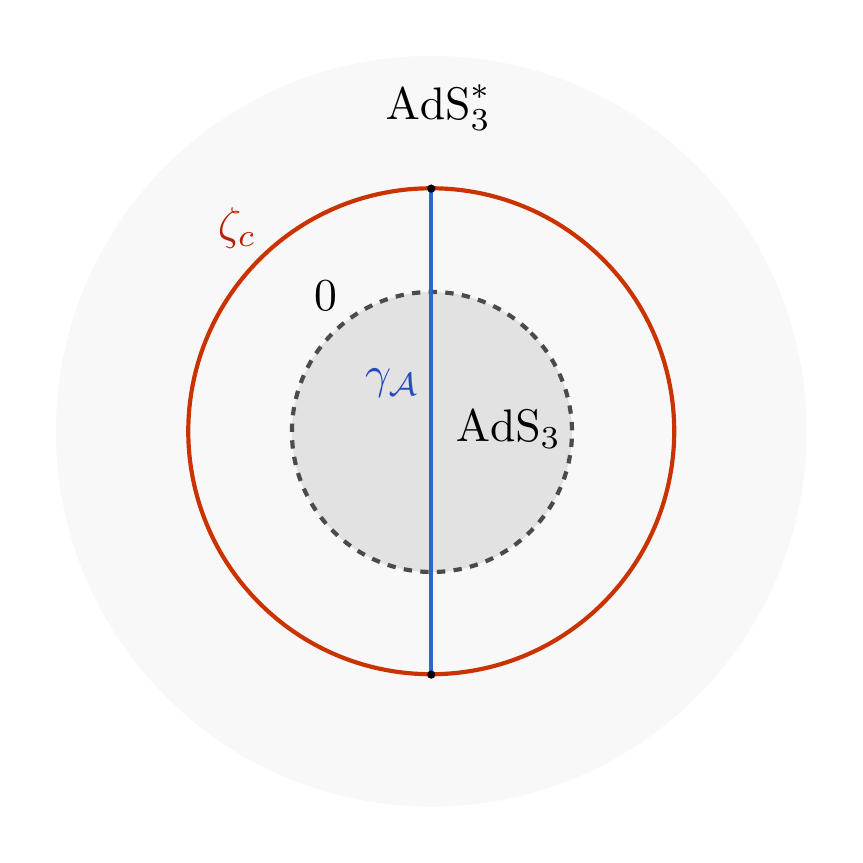}
\end{center}
\caption{A cross section of the glue-on AdS$_3$ space that shows the extended HRT surface (blue) connecting the north and south poles of the sphere at a cutoff surface (red) on AdS$_3^*$. The dashed line denotes the asymptotic boundary of the AdS$_3$ and AdS$_3^*$ regions.}
\label{fig:EAdSRT}
\end{figure}

Let us consider an interval $\mA$ connecting two antipodal points of the sphere at the cutoff $\zeta = \zeta_c < 0$. We would like to find a geometric description of the entanglement entropy \eqref{EEsphere} in the glue-on AdS$_3$ geometry. A subtlety arises due to the minus sign in front of the sphere part of the metric \eqref{glueSphere}. While the interval $\mA$ is spacelike with respect to the boundary metric \eqref{boundarymetricsphere}, it is timelike with respect to the bulk metric \eqref{glueSphere} because $\mA$ resides in the AdS$_3^*$ region. Nevertheless, it is possible to connect the endpoints $\p \mA$ of the interval $\mA$ by an everywhere spacelike surface that goes through the AdS$_3$ part of the space. In fact, there is a natural way to extend the HRT surface from AdS$_3$ to AdS$_3^*$. In order to do so, we first note that the original AdS$_3$ space is a solid ball, and the HRT surface --- which we denote by $\gamma$ --- is just a radial line that starts from $\theta=0$ at the north pole of the asymptotic boundary ($\zeta \to 0^+$), switches to $\theta=\pi$ at the origin ($\zeta = \infty$), and continues to the south pole at $\zeta \to 0^+$. We can extend the HRT surface to the AdS$_3^*$ region of the glue-on AdS$_3$ space \eqref{glueSphere} by extending both ends of the radial line $\gamma$ until they hit the cutoff surface at $\zeta = \zeta_c$. By construction, the two radial half-lines in the AdS$_3^*$ region, which are denoted by $\gamma^*$, are spacelike geodesics.  The glue-on HRT surface is defined as
\eq{
\gamma_\mA= \gamma \cup \gamma^*, \label{se3:sphere_HRT}
} 
which connects the two antipodal points on the cutoff sphere at $\zeta = \zeta_c$ through the glue-on AdS$_3$ space (see  fig.~\ref{fig:EAdSRT} for an illustration). In particular, note that the surface \eqref{se3:sphere_HRT} is everywhere spacelike and piecewise geodesic.

The next step is to assign a geometric invariant to the glue-on HRT surface.  When $\mu<0$, the cutoff surface is located in the interior of AdS$_3$ and $\gamma_\mA$ is a segment of the HRT surface, whose area has been shown to reproduce the entanglement entropy of \TTbar-deformed CFTs \eqref{EEsphere} \cite{Donnelly:2018bef}. More explicitly, we have
\eq{
\label{cutoffHRT} 
\begin{split}
\tilde S[\mA] \equiv \frac{\text{Area}[\gamma_{\mA}]}{4G} & = \frac{\ell}{4G} \int^{\infty}_{\zeta_c} \frac{d\zeta}{\zeta\sqrt{1 + \zeta}}\\
&=\frac{c}{3} \arcsinh \bigg(\sqrt{-\frac{3}{c\mu}}\frac{\ell_\mA}{\pi} \bigg)=S[\mA] ,
\end{split} \qquad \mu<0,
}
where we have used \eqref{dic} together with $c = 3 \ell/2 G$ and $\ell_\mA = \pi \ell$. A natural extension of this expression for the $\mu > 0$ case is to simply extend the lower bound $\zeta_c$ to negative values, with the integrand taking the same form, namely
\eq{
\tilde S[\mA] = {\ell\over 4G} \int^{\infty}_{\zeta_c} \frac{d\zeta}{\zeta\sqrt{1 + \zeta}} = {\ell\over 4G} \left(\int^{\infty}_{\epsilon} \frac{d\zeta}{\zeta\sqrt{1 + \zeta}} +  \int_{\zeta_c}^{-\epsilon} \frac{d\zeta}{\zeta\sqrt{1 + \zeta}}\right), \qquad \mu > 0, \label{glueHRT}
}
where we have introduced a cutoff $\epsilon \to 0^+$ at the asymptotic boundaries of AdS$_3$ ($\zeta = \epsilon$) and AdS$_3^*$ ($\zeta = - \epsilon$). It is not difficult to check that the divergences from the two cutoff surfaces cancel such that the total integral is finite and independent of $\epsilon$. The first term on the right hand side of \eqref{glueHRT} is just the area (length) of the original HRT surface $\gamma \subset \text{AdS}_3$, while the second term is the area (length) of $\gamma^* \subset \text{AdS}_3^*$ multiplied by a minus sign. This motivates us to define the glue-on version of the HRT formula in terms of the signed area 
\eq{\label{HEEEAdS}
\tilde S[\mA] &=\frac{\sArea[\gamma_\mA]}{4G}\equiv \frac{1}{4G}\big({\text{Area}}[\gamma\thinmspace]-{\text{Area}}[\gamma^*] \big).
}
Integrating \eqref{glueHRT}, it is not difficult to verify that the glue-on HRT formula \eqref{HEEEAdS} reproduces the entanglement entropy of \TTbar-deformed CFTs on the sphere \eqref{EEsphere}, namely
\eq{
\tilde S[\mA] =\frac{c}{3} \arccosh \bigg(\sqrt{\frac{3}{c\mu}}\frac{\ell_\mA}{\pi} \bigg)= S[\mA],\qquad \mu>0.  \label{HEEsphere}
} 

The matching \eqref{HEEsphere} suggests that the glue-on version of the HRT formula \eqref{HEEEAdS} can be interpreted as the holographic entanglement entropy of \TTbar-deformed CFTs. In the following, we  provide further support to this interpretation by showing that the glue-on HRT surface $\gamma_\mA$ can be regarded as the minimal surface of the signed area functional 
\eq{
\tilde S[\mA] &= \operatorname*{\min \mop{ext}}
\limits_{\vphantom{\hat{X}}\! X_\mA\sim\mA} \frac{\sArea[X_\mA]}{4G}, \label{minext}
}
where $X_\mA$ is an everywhere spacelike surface in glue-on AdS$_3$ that is homologous to the interval $\mA$ on the cutoff surface on AdS$_3^*$. The signed area of the surface $X_\mA = X \cup X^*$ is the length of the segment lying in the AdS$_3$ region ($X$) minus the length of the segments lying in the AdS$_3^*$ region ($X^*$). The extremality condition implies that the minimal surface has to be piecewise geodesic in both the AdS$_3$ and AdS$_3^*$ regions of the space. We can then prove \eqref{minext} by showing that the radial surface $\gamma_{\mA}$ is indeed extremal by considering infinitesimal variations of the gluing points at the asymptotic boundary. Consider a small deviation from $\gamma_{\mA}$ so that the variation of the tangent vector is parameterized by $\delta\theta'\equiv d\,\delta\theta / d\zeta$ and $\delta\phi'\equiv d\,\delta\phi / d\zeta$. Then the correction to the signed area can be written as  
\eq{
\delta \sArea &=\int_{\zeta_c}^\infty d\zeta\,{1\over\zeta\sqrt{1+\zeta}}\Big(\sqrt{1+\zeta\left((\delta\theta')^2+\sin^2 \theta  \,(\delta\phi')^2\right)}-1\Big)\nonumber\\
&=\int_{\zeta_c}^\infty d\zeta\,{(\delta\theta')^2+\sin^2\theta \,(\delta\phi')^2\over2\sqrt{1+\zeta}} + \dots \label{delta_sArea}
}
where we have ommitted higher order terms in $\delta \theta'$ and $\delta \phi'$. The integrand in \eqref{delta_sArea} is always positive such that the value of the signed area always increases. As a result, the local minimum is given by the radial surface $\gamma_\mA$ which justifies the proposal \eqref{minext}.

We have shown that the glue-on HRT formula \eqref{HEEEAdS} reproduces the entanglement entropy of a half interval of \TTbar-deformed CFTs on the sphere \eqref{EEsphere}. Therefore, it can be identified with the holographic entanglement entropy of the \TTbar-deformed theory. This provides further support for glue-on AdS holography and motivates the more general proposal described in the next section.


\section{Glue-on HRT proposal} \label{se:hrt}

In this section we provide a formal definition of glue-on HRT surfaces and a general prescription for the glue-on HRT formula. A glue-on HRT surface is made of multiple segments that are glued together at the asymptotic boundary. We will show that the extremality condition implies that the vector tangent to the HRT surface must be continuous across the asymptotic boundary, and that its signed area is finite. As an example, we consider in detail the HRT surfaces associated with single and multiple intervals on Poincar\'e AdS$_3$. This analysis reveals the emergence of a minimum length that depends on the $T\bar T$ deformation parameter, and below which no HRT surface exists. We also describe general features of the glue-on HRT formula including positivity,  monotonicity, and subadditivity. 

\subsection{General prescription} \label{se:glueonRTgeneral}

The glue-on HRT formula is proposed to be given by 
\eq{
{\tilde S}[\mA] =
\operatorname*{\min \mop{ext}}
\limits_{\vphantom{\hat{X}}\! X_\mA\sim\mA} \frac{\sArea[X_\mA]}{4G},
\label{proposal}
}
where $\mA$ is an interval at a cutoff surface in the AdS$_3^*$ ($\zeta < 0$) region of a glue-on AdS$_3$ spacetime. The surface $X_\mA$ is homologous to $\mA$ and consists of multiple segments that are glued together at the asymptotic boundary. Let $X_{\epsilon}$ denote the segment in the AdS$_3$ ($\zeta > \epsilon$) region of the spacetime, and $X^*_{\epsilon}$ denote the segment in the AdS$_3^*$ ($\zeta < -\epsilon$) region, where $\epsilon \to 0^+$ is a UV cutoff. The signed area of $X_\mA$ is then given by
\eq{
\sArea[X_\mA] \equiv \lim_{\epsilon\to 0}
\Bigl(
	\mop{Area}[X_{\epsilon}]
	- \mop{Area}[X^*_{\epsilon}]
\Bigr).
\label{eq:signed-area}
}
We conjecture that \eqref{proposal} is a quantity inherently associated with an interval $\mA$ in $T\bar T$-deformed CFTs. In the previous section, we showed that \eqref{proposal} reproduces the entanglement entropy of a half interval on the sphere provided that the UV cutoff is identified with the deformation parameter. More generally, we expect \eqref{proposal} to be related to the entanglement entropy of \TTbar-deformed CFTs in more general scenarios, although its precise relationship to the entanglement entropy is not addressed in this paper.

\bigskip 

\noindent Let us now describe in more detail a few aspects of the proposal \eqref{proposal}.

\paragraph{Finiteness.} The two contributions to the signed area are divergent in the $\epsilon \to 0^+$ limit, but their difference is finite. More explicitly, let us consider the region near the asymptotic boundary where the surface $X_\mA$ crosses from the AdS$_3$ region to the AdS$_3^*$ region of the spacetime. The divergence in the area comes from the $d\zeta^2/4\zeta^2$ term in the metric of any asymptotically AdS$_3$ spacetime \eqref{eq:ads-falloff}. The signed area of the surface $X_\mA$ in this region behaves as
\eq{
\sArea &\sim \int_{\epsilon}^{p}
    \frac{d\zeta}{2\zeta}
+ \int^{-\epsilon}_{p^*}
    \frac{d\zeta}{2\zeta}
= - \frac{1}{2} \log\epsilon+\frac{1}{2} \log \epsilon 
    + \order{1}
= \order{1},
\qquad \epsilon\to 0,
}
where $p > \epsilon$ and $p^* < - \epsilon$ are two points in $X$ and $X^*$. We see that the divergences cancel so that the signed area is finite.

\paragraph{The extremal surface.} The extremization necessarily requires the surface that minimizes the signed area \eqref{proposal}  to be piecewise geodesic. Since $X_\mA$ is required to cross the asymptotic boundary, the signed area should also be extremal under variations of the gluing points. Let us consider a candidate surface $X_\mA$ with multiple segments glued at the asymptotic boundary at the points $\{p_1, \dots, p_{m}\}$. The extremality condition requires the surface $\gamma_\mA$ with the minimum signed area to satisfy 
\eq{
    \delta_{p_i} \sArea[X_\mA]\big|_{X_\mA=\gamma_{\mA}}
    = 0,
\qquad\ i=1,\dots m. \label{eq:extremal-condition}
}
When the boundary is a sphere, we have shown in the previous section that the minimal surface is the analytic extension to AdS$_3^*$ of an HRT surface ending on the asymptotic boundary of AdS$_3$. Generalizations of the HRT formula that consist of different segments, where the minimal surface is obtained by varying the location of the gluing points, have also been considered in other settings. These include the swing surface proposal for the holographic entanglement entropy of asymptotically flat spacetimes \cite{Apolo:2020bld}, as well as in de Sitter holography and timelike entanglement entropy \cite{Doi:2022iyj,Doi:2023zaf,Kawamoto:2023wzj}.

Let us now consider in more detail the extremality condition for the case of a single interval with endpoints $a$ and $b$ parameterized by $(a^i,\zeta_c)$ and $ (b^i,\zeta_c)$. Any candidate extremal surface of \eqref{proposal} necessarily consists of three spacelike geodesics such that 
\eq{
X_{\mA}=X^*_{a,{\hat a}}\cup X_{{\hat a},{\hat b}} \cup X^*_{b,{\hat b}},
}
where $X_{{\hat a},{\hat b}}$ is an AdS$_3$ spacelike geodesic connecting the two gluing points $({\hat a^i},\epsilon)$ and $({\hat b^i},\epsilon)$ at a regulating surface $\mcal{N}_\epsilon$ near the asymptotic boundary of AdS$_3$. On the other hand, $X_{a,{\hat a}}^*$ is a spacelike geodesic in AdS$_3^*$ that connects the endpoint $(a^i, \zeta_c)$ at the cutoff surface to the gluing point $(\hat{a}^i,-\epsilon)$ at a regulating surface $\mathcal N_{-\epsilon}$, and similarly for $X_{b,{\hat b}}^*$. 

Let $D(x^\mu_1,x^\mu_2)$ denote the geodesic distance between any two points $x^\mu_1$ and $x^\mu_2$ such that the signed area of $X_{\mA}$ is given by 
\eq{ \label{se4:D_defs}
\sArea[X_\mA] = D(\hat{a},\hat{b})-D(a,\hat{a})-D(b,\hat b).
}
The extremality condition can then be written as
\eq{
{\p D(\hat{a},\hat{b})\over \p \hat{a}^i }&={\p D(a,\hat{a})\over \p \hat{a}^i },  \qquad {\p D(\hat{a},\hat{b})\over \p \hat{b}^i }={\p D(b,\hat{b})\over \p \hat{b}^i }. \label{extremala}
}
In addition, the fact that $X_{\mA}$ is a spacelike surface implies that the gluing points at the regulating surfaces $\mathcal N_{\pm \epsilon}$ must satisfy the following {\it spacelike condition}: $\hat a$ must be spacelike separated from $a$ while $\hat b$ must be spacelike separated from $b$.

The existence of the extremal surface requires that the points $\hat a$ and $\hat b$ satisfy both the spacelike condition and the extremality condition \eqref{extremala}. As we will show later in explicit examples, when the endpoints $a$ and $b$ are too close to each other, i.e.~when their distance on the cutoff surface is below a scale set by $\sqrt{c\mu}$, the extremality condition \eqref{extremala} cannot be satisfied in the regime where the spacelike condition holds. In this case, there is no extremal surface and we define $\tilde S[\mA] = 0$. This is motivated from the fact that, as we approach the minimum length scale from above, the signed area of the corresponding glue-on HRT surface approaches zero.

Assuming that the extremal surface exists, let us discuss the implications of the extremality condition \eqref{extremala}. First, note that the derivatives of the distance functions with respect to the radial coordinate of the gluing point are given by
\eqsp{
{\p D(\hat{a},\hat{b})\over \p \zeta } &=  -{\ell\over 2\zeta},\qquad \zeta\to 0^+, \\
{\p D(a,\hat{a})\over \p \zeta } = {\p D(b,\hat{b})\over \p \zeta } &= -{\ell\over 2\zeta},\qquad \zeta\to 0^-. 
}
The minus signs are due to the fact that as $|\zeta|$ increases, the distance functions decrease in both the $\zeta > 0$ and $\zeta < 0$ regions. The gradient $\p_\mu D(\hat a, \hat b)$ is normal to the equidistant surface between the points $\hat a$ and $\hat b$. As a result, this covector is tangent to the geodesic connecting $\hat a$ and $\hat b$, and points in the direction that increases the distance to the point~$\hat b$. In fact, Gauss's Lemma (see e.g.~\cite{do1992riemannian}) implies that $\p_\mu D$ is precisely the unit tangent covector, so the normalization is fully determined.

We can parameterize the surface $X_\mA$ locally by $\zeta$, so that it is described by $x^i=x^i(\zeta)$. In general, this parameterization leads to multi-valued functions.\footnote{As an example, consider the usual semicircular spacelike geodesic attached to the asymptotic boundary of $\text{AdS}_3$. When parametrized in terms of $\zeta$, the geodesic goes into the bulk, reaches a maximum value of $\zeta$, and then turns back to the asymptotic boundary. In this case $x^i(\zeta)$ is double valued.} In the following, we focus on a single-valued branch which is always possible near the asymptotic boundary.  Then the tangent vector $\xi^\mu$ of $X_\mA$ is proportional to $({dx^i\over d\zeta},1)$. From the previous discussion, we find that the tangent covector is the gradient of the distance up to a sign, so that near the asymptotic boundaries of AdS$_3$ and AdS$_3^*$ we have
\renewcommand{\arraystretch}{2}
\eqsp{
\xi_\mu = -\ell\bigg(2\gamma_{ij}{dx^j\over d\zeta},{1\over 2\zeta}\bigg) \bigg|_{\zeta\to 0^\pm} &= \left\{\begin{array}{c}
\bigg(\dfrac{\p D(\hat{a},\hat{b})}{ \p \hat{a}^i },\,\dfrac{\p D(\hat{a},\hat{b})}{ \p \zeta }\bigg),\qquad \zeta\to 0^+, \\
\bigg(\dfrac{\p D(a,\hat{a})}{ \p \hat{a}^i },\,\dfrac{\p D(a,\hat{a})}{ \p \zeta }\bigg),\qquad \zeta\to 0^-,
\end{array}\right.
\label{eq:gradient-tangent}
} 
where the normalization is fixed by matching the radial components. The extremality condition \eqref{extremala} then implies that the tangent vector is continuous across the asymptotic boundary. More explicitly, since the glue-on construction guarantees that the two-dimensional metric $\gamma_{ij}$ is continuous across the asymptotic boundary, the derivatives ${dx^i\over d\zeta}$ should also be continuous, such that
\eq{
{dx^i\over d\zeta}\bigg|_{\zeta\to0^+}={dx^i\over d\zeta}\bigg|_{\zeta\to0^-}.
\label{eq:curve-continuity}
}
Note that \eqref{eq:curve-continuity} depends crucially on the asymptotic behavior of AdS$_3$ and AdS$_3^*$, as well as on the signed area. 

Recall that the extremal surface is piecewise geodesic in both the AdS$_3$ and AdS$_3^*$ regions of the spacetime. The condition \eqref{eq:curve-continuity} at the gluing point tells us that the geodesic in the AdS$_3^*$ region can be constructed from that in AdS$_3$ by simply continuing the range of the radial coordinate $\zeta$ to negative values. More explicitly, let us parameterize the single-valued branch of the HRT surface in AdS$_3$ that is attached to the point $\hat a$ at the asymptotic boundary by
\eq{
\gamma_{\hat a}: \quad x^i = f^i(\zeta), \qquad \epsilon < \zeta \le \zeta_{\text{max}}, \qquad f^i(\epsilon) = \hat a^i,
} 
where $\zeta_{\text{max}}$ is the turning point of the HRT surface, i.e.~the maximum value of $\zeta$. Then, its extension to the entire glue-on AdS$_3$ spacetime is given by
\eq{
\gamma_{a}:\quad x^i = f^i(\zeta), \qquad \zeta_c \le \zeta \le \zeta_{\text{max}}, \qquad f^i(\zeta_c) = {a}^i,
}
such that $\gamma_{a}$ satisfies the continuity condition \eqref{eq:curve-continuity} and ends on the point $a$ at the cutoff surface. Since the metric of AdS$_3^*$ is obtained by a similar extension, the part of $\gamma_{a}$ lying in the AdS$_3^*$ region is automatically geodesic. Similarly, we can continue the other single-valued branch $\gamma_{\hat b}$ of the HRT surface attached to $\hat b$ in AdS$_3$ into a surface $\gamma_b$ that attaches to the point $b$ at the cutoff surface in AdS$_3$. The full glue-on HRT surface is then given by
\eq{
\gamma_{\mA} = \gamma_{a} \cup \gamma_{b}.\label{multi}
}

The above argument can be generalized to multiple intervals in a straightforward way. As we are considering intervals in a two-dimensional cutoff surface, the boundary of any interval $\mA$ consists of an even number of points, which can be grouped into pairs as $\{a_{(k)},b_{(k)}\}$ with $k=1,\dots, n$. For a given grouping, the extremal surface will then be the union $\cup_k \gamma_{(k)}$, where $\gamma_{(k)}$ is the glue-on HRT surface anchored at $\{a_{(k)},b_{(k)}\}$. There are several ways of grouping the endpoints into pairs, and the final glue-on HRT surface is the one that minimizes the value of the signed area,
\eq{
\tilde{S}[\mA]=\min_{\textit{pairings}\,} {\sArea [\cup_{k}\gamma_{(k)}]\over 4G}.
}


\subsection{Single interval in Poincar\'e AdS$_3$} \label{se:ERTPoincare}

In this section we illustrate in detail how the general prescription \eqref{proposal} works for single intervals in the glue-on version of Poincar\'e AdS$_3$. In particular, we discuss the emergence of a minimum length that depends on the \TTbar deformation parameter and below which the glue-on HRT surface ceases to exist.

Let us consider the $T\bar T$ deformation of the vacuum of a CFT on the plane. The theory is proposed to be dual to the cutoff/glue-on version of Poincar\'e AdS$_3$ 
\eq{
 \quad d s^{2}=\ell^{2}\left(\frac{d \zeta^{2}}{4 \zeta^{2}} + \frac{d w^+ d w^-}{\zeta}\right), \qquad \zeta \ge\zeta_c, \label{poincare2}
}
where the cutoff surface is located at $\zeta = \zeta_c = -c \mu/3\ell^2$. Depending on the sign of $\mu$, the cutoff surface may be located inside AdS$_3$ ($\zeta > 0$) or in the interior of the AdS$_3^*$ spacetime ($\zeta < 0$). In both cases, the background metric the deformed theory couples to can be read from the line element $ds_c^2$ at the cutoff surface such that $ds^2 |_{\zeta_c} = \zeta_c^{-1} ds_c^2$ where $ds_c^2 = \ell^2 dw^+ dw^-$. Since the coordinates $w^\pm = x \pm t$ are not compactified, the corresponding $T\bar T$-deformed CFT is defined on the plane.

\begin{figure}[!ht]
{\centering
\subfloat[$\ell_\mA > \ell_{\min}$ \label{fig:poincareRTa}]{
 \includegraphics[scale=0.478]{img/RTPoincare3}
}\hfill
\subfloat[$\ell_\mA = \ell_{\min}$\label{fig:poincareRTb}]{
 \includegraphics[scale=0.478]{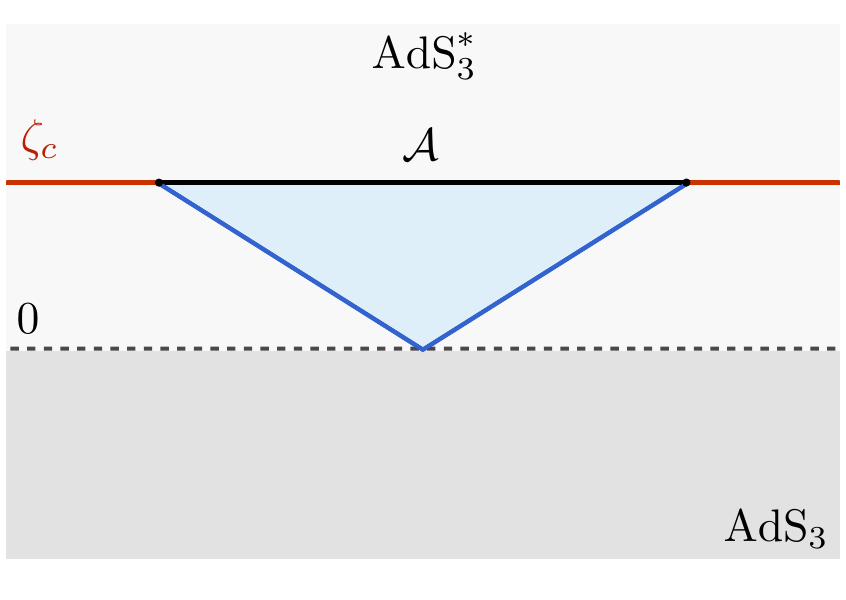}
}
}
\caption{Fixed-time slices of the glue-on Poincar\'e spacetime showing two HRT surfaces (blue) associated with an interval $\mA$ at a finite cutoff (red) in AdS$_3^*$. When the length of the interval $\ell_\mA$ equals $ \ell_{\min}$, the AdS$_3$ part of the HRT surface shrinks to a point and the AdS$_3^*$ part becomes lightlike.}
\label{fig:poincareRT}
\end{figure}

We now focus on the $\mu > 0$ case and consider a spacelike interval $\mA$ on the cutoff surface. In terms of the dimensionful coordinates $(\ell w^+, \ell w^-)$, the endpoints $\p \mA$ can be parametrized by
\eq{
\p\mA = \bigg\{ \biggl(-\frac{ \ell^+}{2} , -\frac{  \ell^-}{2} \biggr), \bigg(\,\frac{  \ell^+}{2} , \frac{  \ell^-}{2}  \bigg)\bigg\}, \qquad \ell^+ \ell^- > 0, \label{intervalPoincare}
}
such that the total length of the interval is $\ell_\mA = \sqrt{\ell^+ \ell^-}$. The requirement $ \ell^+ \ell^- > 0$ guarantees that $\mA$ is spacelike with respect to the line element $ds_c^2$ at the cutoff surface. We can obtain the glue-on HRT surface by extending a spacelike geodesic in AdS$_3$ across the asymptotic boundary towards the cutoff surface in the AdS$_3^*$ region. This leads to
\eq{\renewcommand{\arraystretch}{.5}
\gamma_{\mA}: \,	\left\{ \begin{array}{l} 
	\dfrac{w^+}{\ell^+} = \dfrac{w^-}{\ell^-}, \\
	\\
	\zeta - \zeta_c + w^+ w^- = \dfrac{\ell^+ \ell^-}{4\ell^2},
\end{array} \right.  \qquad \zeta \ge \zeta_c . \label{RTcutoff2}
}
The equations describing the glue-on HRT surface $\gamma_\mA$ \eqref{RTcutoff2} take the same form as those describing the HRT surface in pure $\text{AdS}_3$ except that $\zeta$ can now be negative. As illustrated in fig.~\ref{fig:poincareRTa}, $\gamma_\mA$ is made of two parts that are glued at the asymptotic boundary,
\eq{
	\gamma_{\mA} = \gamma\cup\gamma^*.
} 
The first part of the glue-on HRT surface $\gamma_\mA$ is denoted by $\gamma^*$ and consists of two hyperbolic segments that lie on the AdS$_3^*$ region of the spacetime. These segments are attached to the endpoints of the interval $\mA$ at the cutoff surface and extend towards the asymptotic boundary at $\zeta = 0^-$, where they attach to the endpoints $\p\hat\mA$ of an auxiliary interval $\hat\mA$ parametrized by
\eq{
\p\hat\mA = \bigg\{ \biggl(-\frac{ \hat\ell^+}{2} , -\frac{  \hat\ell^-}{2} \biggr), \bigg(\,\frac{  \hat\ell^+}{2} , \frac{ \hat \ell^-}{2}  \bigg)\bigg\}, \qquad \zeta = 0. \label{poincareauxint}
}
The auxiliary variables $\hat \ell^\pm$ are related to the physical ones $\ell^\pm$ by
\eq{
\hat{\ell}^\pm = \ell^\pm \sqrt{1 + \frac{4\zeta_c}{\ell^+ \ell^-}}.
\label{auxlengths}
}
Since the AdS$_3^*$ spacetime is also Poincar\'e, any spacelike surface in AdS$_3^*$ connecting the endpoints of $\mA$ must necessarily cross to the AdS$_3$ region through the asymptotic boundary. 

The second part of the glue-on HRT surface, which is denoted by $\gamma$, lies in the $\zeta \ge 0$ region of the spacetime. It consists of a semicircle (the standard HRT surface) that is attached smoothly to the hyperbolic segments $\gamma^*$ at the endpoints $\p\hat\mA$ at the asymptotic boundary. In order for the surface $\gamma_{\mA}$ to extend to the AdS$_3$ ($\zeta > 0$) region of the spacetime, the auxiliary parameters ${\hat \ell}^\pm $ in \eqref{auxlengths} must be real and nonvanishing. This requirement constraints $\mA$ to be larger than a minimum value such that
\eq{
\ell_{\mA} >\ell_{\min}\equiv 2\ell \sqrt{-\zeta_c }=2\sqrt{c\mu\over3}.
\label{poincareminlength}
}
The emergence of a minimum length for the interval $\mA$ is consistent with our expectations from \TTbar, studies of which suggest that physically meaningful distances should be larger than the scale of nonlocality of the theory \cite{Dubovsky:2012wk}. The latter is proportional to the square root of the deformation parameter as in \eqref{poincareminlength}. 

\paragraph{The extremality condition.}

We will now show that $\gamma_\mA$ is indeed the extremal surface of minimum signed area when $\ell_\mA > \ell_{\min}$. For simplicity, let us consider an interval on the $t = 0$ slice of the cutoff surface such that $\ell^+ = \ell^- = \ell_\mA$. We consider a two-parameter family of surfaces $X_\mA = X^{*}_{a, \hat a}\cup X_{{\hat a},{\hat b}} \cup X^*_{b, \hat b}$ that consist of three segments described by
\eq{
X_\mA\colon \, \left\lbrace
\begin{aligned}\ \,
X^*_{a, \hat{a}}\colon\ \,&
	(\hat{a} - a)\,\zeta
    + (x - \hat{a}) \bigl(
        (x - a)(\hat{a} - a) + \zeta_c
    \bigr) = 0, \\[.4ex]
\ X_{\hat{a},\hat{b}}\colon\ \,&
	\zeta + (x - \hat{a})(x - \hat{b}) = 0,\\[.4ex]
X^*_{b, \hat{b}}\colon\ \,&
	(\hat{b} - b)\,\zeta
    + (x - \hat{b}) \bigl(
        (x - b)(\hat{b} - b) + \zeta_c
    \bigr) = 0, \\[.4ex]
\end{aligned}
\right.\label{eq:perturbed-HRT}
}
where $a = -{\ell_\mA}/{2\ell}$ and $b = {\ell_\mA}/{2\ell}$. The $X^*_{a,\hat{a}}$ and $X^*_{b,\hat{b}}$ segments are required to be spacelike surfaces connected to the left and right endpoints of the interval, respectively. This leads to the following range for $\hat a$ and $\hat b$
\eq{
	-\frac{\ell_\mA + \ell_{\min}}{2\ell} < \hat{a} < -\frac{\ell_\mA - \ell_{\min}}{2\ell},\qquad
 \frac{\ell_\mA - \ell_{\min}}{2\ell} < \hat{b} &< \frac{\ell_\mA + \ell_{\min}}{2\ell}, \label{eq:ab-params-range}
}
where we have used the definition of $\ell_{\min}$ in \eqref{poincareminlength}. These bounds are saturated when the spacelike surfaces $X^*_{a,\hat{a}}$ and $X^*_{b,\hat{b}}$ approach the lightcone of the endpoints. Note that the surfaces $X_\mA$ exist for any $\ell_\mA > 0$ but are generically not extremal. It is straightforward to verify that $(i)$ the three segments making up $X_\mA$ are all spacelike geodesics such that the surface $X_\mA$ is everywhere extremal except, generically, at the gluing points; and $(ii)$ these segments are glued at the asymptotic boundary and anchored at $\pd\mA$, such that the piecewise geodesic $X_\mA$ is homologous to $\mcal{A}$. 

The signed area of the $X_\mA$ surfaces can be written as in \eqref{se4:D_defs} where the distance functions are explicitly given by
\eq{
D(\hat a, \hat b)=\ell \log \bigg( {(\hat a-\hat b)^2\over \epsilon }\bigg) ,\qquad  
D( a, \hat a)={\ell} \log \biggl(-{ (a - \hat{a})^2 +\zeta_c \over \sqrt{-\zeta_c \epsilon}}\biggr),
}
where $\epsilon \to 0^+$ regulates the location of the asymptotic boundaries of AdS$_3$ and AdS$_3^*$. The distance function $D(b, \hat b)$ can be obtained from $D(a, \hat a)$ by letting $(a, \hat a) \leftrightarrow (b,\hat b)$. As described earlier, the signed area is finite and independent of the regulator. The extremality condition \eqref{extremala} then implies that
\eq{
0   &= 2\ell\,\frac{
		(\ell_\mA/\ell - 2\hat{a})
        (\ell_\mA/\ell - 2\hat{b})
		+4\zeta_c
	}{
		(\hat{b} - \hat{a})
		\big((\ell_\mA/\ell - 2\hat{b})^2 + 4\zeta_c\big)
	}, \qquad 0 = 2\ell\,\frac{
		(\ell_\mA/\ell + 2\hat{a})
        (\ell_\mA/\ell + 2\hat{b})
		+4\zeta_c
	}{
		(\hat{a} - \hat{b})
		\big((\ell_\mA/\ell + 2\hat{a})^2 + 4\zeta_c\big)
	}. \label{eq:extremal-condition-poincare}
}
When $\ell_\mA > \ell_{\min}$, the extremality condition \eqref{eq:extremal-condition-poincare} is satisfied if, and only if,
\eq{
	-\hat{a} = \hat{b}
 = \frac{1}{2\ell} \sqrt{\ell_\mA^2 - \ell^2_{\min}}
 = \frac{1}{2\ell} \sqrt{\ell_\mA^2 + 4\ell^2\zeta_c},
\label{eq:extremal-solution}
}
which are precisely the endpoints of the auxiliary interval $\pd\hat\mA$ defined in \eqref{poincareauxint} and \eqref{auxlengths}. The surface $X_\mA$ satisfying \eqref{eq:extremal-solution} is nothing but the extremal surface $\gamma_\mA$. Furthermore, a second order variation around the extremal point \eqref{eq:extremal-solution} shows that it indeed corresponds to the minimum signed area, namely
\eq{
\sArea[X_\mA] \ge
\sArea[\gamma_\mA]
= 2\ell \arccosh\biggl(
	\frac{\ell_\mA}{2\ell\sqrt{-\zeta_c}}
\,\biggr).
\label{eq:minimalArea}
}
When $\ell_\mA < \ell_{\min}$, the solution \eqref{eq:extremal-solution} becomes imaginary and there is no real solution to the extremality condition \eqref{eq:extremal-condition-poincare}, so the signed area $\sArea[X_\mA]$ has no extremal point.

\begin{figure}[!t]
\centering
\includegraphics[scale=0.478]{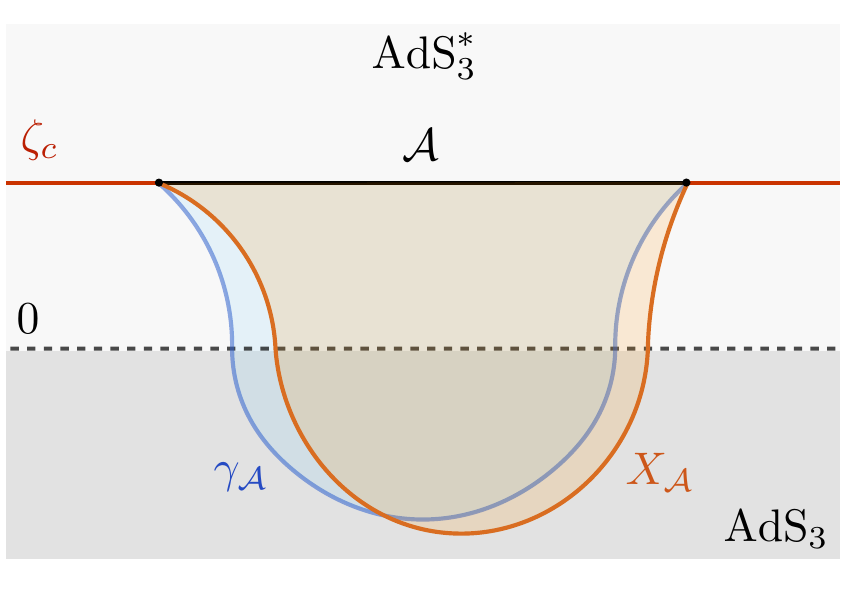}
\caption{A non-extremal surface $X_\mA$ (orange) and the extremal surface $\gamma_\mA$ (blue). Both surfaces are piecewise geodesic and anchored at the endpoints of the interval $\mcal{A}$ on the cutoff surface (red) at $\zeta = \zeta_c$.
}
\end{figure}

In addition, we note that for generic values of $\hat a$ and $\hat b$, the surface $X_\mA$ is not smooth around these points. Indeed, the tangent vector along $X_\mA$ is discontinuous across the asymptotic boundaries of AdS$_3$ and AdS$_3^*$, namely between the $\zeta \to 0^+$ and $\zeta \to 0^-$ surfaces of the glue-on AdS$_3$ spacetime. Using \eqref{eq:gradient-tangent}, the discontinuity in the tangent vector is related to the variation of the signed area
\eq{
	\Big(\,
    \frac{dx}{d\zeta}  \Big|_{\,\zeta\to 0^+}
	-
	\frac{dx}{d\zeta}  \Big|_{\,\zeta\to 0^-}
    \Big)_{x = \hat a}
    &= -\frac{1}{2\ell}\,\pd_{\hat{a}} \sArea[X_\mA],\\[1ex]
	\Big(\,
    \frac{dx}{d\zeta}  \Big|_{\,\zeta\to 0^+}
	-
	\frac{dx}{d\zeta}  \Big|_{\,\zeta\to 0^-}
    \Big)_{x = \hat b}
    &= -\frac{1}{2\ell}\,\pd_{\hat{b}} \sArea[X_\mA].
}
We have thus verified that the extremality condition \eqref{eq:extremal-condition-poincare} implies an identification of the first derivatives, namely \eqref{eq:curve-continuity}. This provides the justification for the analytic continuation of the glue-on HRT surface \eqref{RTcutoff2}: if we start from the continuity condition \eqref{eq:curve-continuity}, and consider the geodesic equations on both sides of the asymptotic boundary, we end up with the unique analytic solution \eqref{RTcutoff2}.

\paragraph{The glue-on HRT formula.}
Using \eqref{eq:minimalArea} and the dictionary \eqref{dictionary} we obtain
\eq{
\tilde S[\mA] = \frac{\sArea[\gamma_\mA]}{4G} =	\left\lbrace
\begin{aligned}
&\frac{c}{3} \arccosh\biggl( \sqrt{\frac{3}{\smash[b]{ c\mu }}} \frac{\ell_\mA}{2} \thinmspace  \biggr),
& \qquad &\mu > 0,
	\\[.5ex]
&\frac{c}{3} \arcsinh\biggl( \sqrt{-\frac{3}{\smash[b]{ c\mu }}} \frac{\ell_\mA}{2} 
 \thinmspace \biggr),&\qquad &\mu<0 ,
\end{aligned}
 \right. \label{HEEpoincareAdS}
}
where we have also included the $\mu < 0$ result previously obtained in~\cite{Lewkowycz:2019xse}. By construction, the first line of \eqref{HEEpoincareAdS} is only valid when the interval is larger than the minimum length, $\ell_\mA > \ell_{\min}$. The limiting case where $\ell_\mA = \ell_{\min}$ corresponds to the case when the glue-on HRT surface $\gamma_{\mA}$ approaches a lightlike geodesic, and thus the signed area approaches zero. When $\ell_\mA < \ell_{\min}$, there is no everywhere spacelike curve that extremizes the signed area, and hence the glue-on HRT surface ceases to exist. In this case, we have defined $\tilde S[\mA]=0$. 
 
It is interesting to note that the peculiar behavior of $\tilde S[\mA]$ observed above is also found in the entanglement entropy of the undeformed CFT when the size of the interval $\ell_\mA$ becomes less than or equal to the UV cutoff $\epsilon$. Indeed, we observe that when $\ell_\mA = \epsilon$, the entanglement entropy $S_{\textrm{CFT}}[\mA] = (c/3)\log(\ell_\mA/\epsilon)$ vanishes, and that it becomes negative when $\ell_\mA < \epsilon$. This is not surprising as the entanglement entropy of an interval whose size is smaller than the UV cutoff is not physically well defined. Consequently, our results are consistent with the fact that, although UV-complete, $T\bar T$-deformed CFTs feature a minimum length that is proportional to the square root of the deformation parameter~\cite{Dubovsky:2012wk}. This suggests that there is a close relationship between the glue-on HRT formula and the entanglement entropy of \TTbar-deformed CFTs beyond the case of a half interval on the sphere considered in section \ref{se:ERTsphere}. In particular, note that a minimum length of the interval has also been observed for the HRT surface in the single-trace version of the $T\bar T$ deformation with a positive deformation parameter~\cite{Chakraborty:2018kpr}.


\subsection{Multiple intervals and phase transitions}
\label{se:multi-intervals}

\begin{figure}[!ht]
{\centering
\subfloat[$\ell_x > \ell_{\min}$ \label{fig:multiRTa}]{
 \includegraphics[scale=0.478]{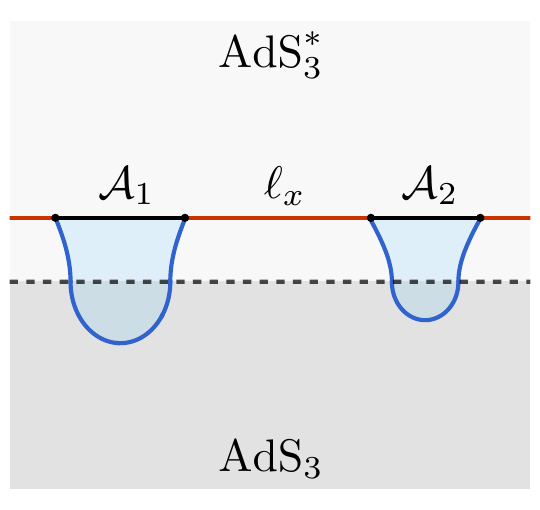}
}\hfill
\subfloat[$\ell_x > \ell_{\min}$\label{fig:multiRTb}]{
 \includegraphics[scale=0.478]{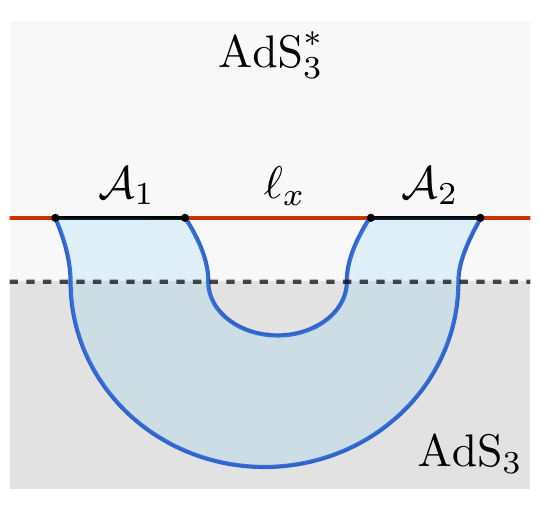}
}\hfill
\subfloat[$\ell_x < \ell_{\min}$\label{fig:multiRTc}]{
 \includegraphics[scale=0.478]{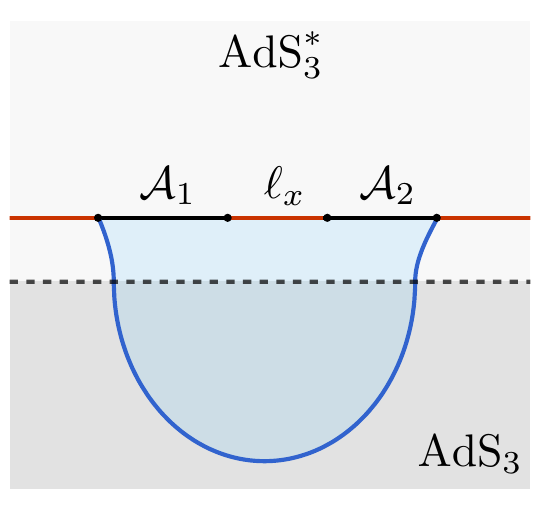}
}
}
\caption{The glue-on HRT surfaces (blue) associated with two intervals $\mA_1$ and $\mA_2$ on a fixed-time slice at a cutoff (red) in the glue-on version of Poincar\'e AdS$_3$. Cases (a) and (b) show the two competing surfaces that are possible when the separation $\ell_x$ between the intervals is greater than $\ell_{\min}$. When $\ell_x < \ell_{\min}$, there is no glue-on HRT surface associated with $\ell_x$ and the disjoint intervals $\mA_1$ and $\mA_2$ are treated as one, as illustrated in (c).
}
\label{fig:multiRT}
\end{figure}

The existence of a minimum length \eqref{poincareminlength} leads to interesting consequences for the glue-on HRT surfaces of disjoint intervals. In order to illustrate this, let us consider two intervals $\mA_1$ and $\mA_2$ of sizes $\ell_{1}$ and $\ell_{2}$, respectively, that are separated by some distance $\ell_x$ on the same fixed-time slice (see fig.~\ref{fig:multiRT}). The disjoint interval $\mA_1 \cup \mA_2$ has four endpoints which can be grouped into two pairs in two different ways. Assuming that the separation between the intervals is greater than the minimum length \eqref{poincareminlength}, we find that the glue-on HRT formula \eqref{multi} reads
\eq{
{\tilde S}[{\mA_1\cup \mA_2}] &= \min\bigl\{ \tilde S_{\ell_{1}} +  \tilde S_{\ell_{2}}, \, \tilde S_{\ell_1 + \ell_2 + \ell_x} + \tilde S_{\ell_x} \bigr\}, \qquad \ell_1,\ell_2,\ell_x \ge \ell_{\min},
\label{multipleint}
}
where $\tilde S_{\ell_\mA}$ is defined for convenience by
\eq{
\tilde S_{\ell_\mA} \equiv \frac{c}{3} \arccosh \Big(\frac{\ell_\mA}{\ell_{\min}}\Big). \label{eq:s-function}
}
The first term in \eqref{multipleint} comes from two disconnected glue-on HRT surfaces, as shown in fig.~\ref{fig:multiRTa}, while the second term comes from the connected contribution shown in fig.~\ref{fig:multiRTb}.

When the sizes of the intervals are sufficiently close to $\ell_{\min}$, it is possible to show that only the disconnected HRT surface dominates when $\ell_x \ge \ell_{\min}$. More precisely, we find that $\tilde S_{\ell_{1}} + \tilde S_{\ell_{2}} \le \tilde S_{\ell_1 + \ell_2 + \ell_x} + \tilde S_{\ell_x}$ for any $\ell_x$ provided that $\ell_1 + \ell_2 \le (1 + \sqrt{5})\, \ell_{\min}$, namely
\eq{
\tilde S[\mA_1\cup \mA_2]  =  \tilde S_{\ell_{1}} + \tilde S_{\ell_{2}}, \qquad 2\ell_{\min} < \ell_1 + \ell_2 \le (1 + \sqrt{5})\, \ell_{\min}, \qquad  \ell_x \ge \ell_{\min}.
}
Interestingly, a similar result can be obtained from the holographic entanglement entropy in AdS$_3$/CFT$_2$ if we identify $\ell_{\min}$ with the size of the UV cutoff $\epsilon$.

\begin{figure}[ht!]
\centering
\mbox{  
\parbox[c]{.4\textwidth}{
\subfloat[
	$\ell_1 + \ell_2 < (1 + \sqrt{5})\,\ell_{\min}$
	\label{fig:TwoIntervalsTwoPhases}
]{
	\includegraphics[scale=1]{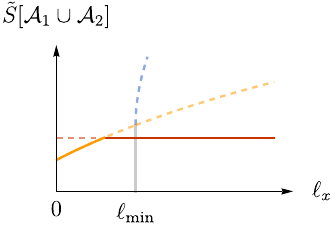}
}
}
\parbox[c]{.2\textwidth}{
\hspace*{-1em}
\vspace*{2ex}
	\includegraphics[scale=1]{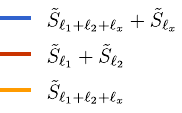}
}
\hspace*{-1.5em}
\parbox[c]{.4\textwidth}{
\subfloat[
	$\ell_1 + \ell_2 > (1 + \sqrt{5})\,\ell_{\min}$%
	\label{fig:TwoIntervalsThreePhases}
]{
	\includegraphics[scale=1]{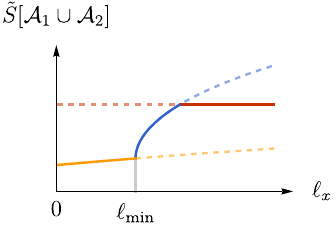}
}
}
}
\caption{$\tilde{S}[\mA_1\cup\mA_2]$ of two intervals $\mcal{A}_1$ and $\mcal{A}_2$ of sizes $\ell_1$ and $\ell_2$, as a function of the distance $\ell_x$ in between. When $\ell_x \lesssim \ell_{\min}$ the dominant phase is always the single interval configuration, as shown fig.~\ref{fig:multiRTc}.
As shown in (a), for sufficiently small $\ell_1$ and $\ell_2$, the ``bridge'' HRT surface (fig.~\ref{fig:multiRTb}), whose entropy is given by the blue curves above, never dominates.
}
\label{fig:phases-plot}
\end{figure}

Another novelty arises when $\ell_x < \ell_{\min}$. In this case, the smaller HRT surface in fig.~\ref{fig:multiRTb} ceases to exist. This means that although the intervals $\mA_1$ and $\mA_2$ are separated by a distance $\ell_x$, they may be effectively treated as a single joint interval.  The HRT surfaces of the single and multiple intervals are allowed to compete and the transition point is determined dynamically, by minimizing between $\tilde S_{\ell_1} + \tilde S_{\ell_2}$ and $\tilde S_{\ell_1 + \ell_2 + \ell_x}$ such that
\eq{
\tilde S[\mA_1\cup \mA_2]  = \min \bigl\{ \tilde S_{\ell_1} + \tilde S_{\ell_2},\, \tilde S_{\ell_1 + \ell_2 + \ell_x}  \bigr\}, \qquad \ell_x < \ell_{\min}. \label{multipleint2}
}
The transitions of $\tilde{S}[\mA_1 \cup \mA_2]$ are illustrated in fig.~\ref{fig:phases-plot}. 

It is interesting to note that the behavior \eqref{multipleint2} of the glue-on HRT formula is similar to the holographic entanglement entropy of the undeformed CFT when $\ell_x$ is smaller than the UV cutoff $\epsilon$ of the theory. In this case, any separation $\ell_x$ smaller than $\epsilon$ is unphysical, so the disjoint intervals $\mA_1$ and $\mA_2$ can behave as a single one. Relatedly, since the glue-on HRT surface cannot resolve subregions of size $ \ell_{\mA} < \ell_{\min}$, the extremal surface associated with a region with multiple holes of size $\ell_x < \ell_{\min}$ cannot be distinguished from that of a region without any holes. This is compatible with the interpretation that $\ell_{\min}$ corresponds to a minimum distance in the dual field theory.

To summarize, we have found that the glue-on HRT formula for two disjoint intervals in $T\bar T$-deformed CFTs with $\mu > 0$ is given by
\eq{
\tilde S[\mA_1\cup \mA_2]  = \left\{
\begin{aligned}
	& \min \bigl\{\thinmspace
			\tilde S_{\ell_1} + \tilde S_{\ell_2},\, \tilde S_{\ell_1 + \ell_2 + \ell_x} + \tilde S_{\ell_x} \bigr\}, \quad&\ell_1,\ell_2 \ge \ell_{\min},   & & \ell_x \ge \ell_{\min}, \\
	& \min \bigl\{\thinmspace
			\tilde S_{\ell_1} + \tilde S_{\ell_2},\, \tilde S_{\ell_1 + \ell_2 + \ell_x}  \bigr\}, &\ell_1,\ell_2 \ge \ell_{\min}, && \ell_x < \ell_{\min}.
\end{aligned}
\right. \label{eq:EEtwoIntervals}
}
where $\tilde S_{\ell_i}$ is defined in \eqref{eq:s-function} and $\ell_{\min} =2\sqrt{c\mu/3}$.


\subsection{Features of the glue-on HRT formula in Poincar\'e AdS$_3$} \label{se4:featuresHRT}

Let us now describe some interesting properties of the glue-on HRT formula for a single interval on Poincar\'e AdS$_3$ \eqref{HEEpoincareAdS} with $\mu > 0$:

\begin{itemize}
\item \emph{Positivity.} Although the signed area contains a minus sign, the glue-on HRT formula always yields non-negative results in Poincar\'e AdS$_3$. This can be verified directly from the HRT formula for a single interval \eqref{HEEpoincareAdS}. The multi-interval result is also non-negative since it is built from the single-interval HRT formula and it does not introduce any additional minus signs. Furthermore, note that other locally AdS$_3$ spacetimes can be obtained from Poincar\'e AdS$_3$ by a coordinate transformation which leaves the geodesic distance invariant. This suggests that positivity of the glue-on HRT formula also holds for other glue-on AdS$_3$ spacetimes in Einstein gravity. However, note that in other spacetimes, there might be other extremal surfaces that cannot be obtained from a coordinate transformation and are the result of a non-trivial topology. We will come back to this point later when we consider glue-on versions of global AdS$_3$ and the BTZ black hole.     

\item \emph{Purity}. An interval $\mA$ shares the same endpoints with its complement $\mA^c$. Since Poincar\'e AdS$_3$ has a trivial topology, the glue-on version of the HRT surface is the same for both $\mA$ and $\mA^c$, such that
\eq{
\tilde S[\mA]=\tilde S[\mA^c].
}
This is similar to the fact that the entanglement entropy of an interval on a pure state is the same as that of its complement. For this reason we refer to this property as \emph{purity}.

\item \emph{$C$-function. } In analogy with the Casini-Huerta $C$-function \cite{Casini:2012ei}, we can define
\eq{
C\equiv\ell_\mA\p_{\ell_\mA} \tilde S[\mA].
} 
For Poincar\'e AdS$_3$ we then have 
\eq{
C={c\over 3\sqrt{1- { 4c\mu\over 3\ell_\mA^2}}},
}
which is positive for all $\mu$ provided that $\ell_\mA > \ell_{\min}$. This agrees with the $C$-function obtained from the holographic entanglement entropy for \TTbar deformed CFTs in the $\mu<0$ case computed in \cite{Lewkowycz:2019xse}.

\item \emph{Monotonicity.} 
Positivity of the $C$-function guarantees that $\tilde S[\mA]$ is monotonic as long as the length of the interval is larger than or equal to $\ell_{\min}$, namely,
\eq{\tilde S[\mA_1]>\tilde S[\mA_2],\qquad  \forall\ \, \ell_1>\ell_2\ge \ell_{\min}, 
}
where $\ell_1$ and $\ell_2$ are the lengths of the intervals $\mA_1$ and $\mA_2$, respectively.

\item \emph{Concavity.}
It is not difficult to check that $\tilde S[\mA]$ is also concave in the range $\ell_{\mA}>\ell_{\min}$, 
namely
\eq{
\p^2_{\ell_\mA}\tilde S[\mA]
= -\frac{c}{3} \frac{\ell_\mA}{\,(\ell_\mA^2 - \ell_{\min}^2)^{3/2}}<0,
\qquad \forall\ \, \ell_\mA > \ell_{\min}.\label{eq:concavity}
}

\item \emph{Subadditivity.}
Consider two adjacent intervals $\mA_1$ and $\mA_2$ of lengths $\ell_1$ and $\ell_2$. In general, subadditivity is violated when the following function is negative
\eq{ I_2\equiv \tilde S[\mA_1]+\tilde S[\mA_2]-\tilde S[\mA_1 \cup \mA_2].
}
When either $\ell_1\le \ell_{\min}$ or $\ell_2 \le \ell_{\min}$, then $I_2<0$ and subadditivity is violated. This follows from the monotonicity of $\tilde S[\mA]$ and the fact that either $\tilde S[\mA_1]$ or $\tilde S[\mA_2]$ vanish. On the other hand, when $\ell_1,\,\ell_2> \ell_{\min}$, we have
\eq{
I_2 = \tilde S_{\ell_1}+\tilde S_{\ell_2}-\tilde S_{\ell_1+\ell_2},
}
where $\tilde S_{\ell_i}$ is the function defined in \eqref{eq:s-function}. We note that given $\ell_2$, the function $I_2$ is a monotonic function of $\ell_1$, and for $\ell_1$ sufficiently close to the minimum length, $I_2$ is always negative. When $\ell_1=\ell_2$, we find that the zero of $I_2$ is located at $\ell_1=\ell_2= {1\over 2}(1+\sqrt{3})\,\ell_{\min}$. This leads to a sufficient condition for subadditivity to be satisfied, namely,
\eq{
I_2>0 \quad \text{for} \quad \ell_1,\,\ell_2>{1\over 2}(1+\sqrt{3})\,\ell_{\min},
}
and similarly, it leads to a sufficient condition for subadditivity to be violated,
\eq{
I_2<0 \quad \text{for} \quad \ell_1,\,\ell_2<{1\over 2}(1+\sqrt{3})\,\ell_{\min}. 
}
If we fix $\ell_2>{1\over 2}(1+\sqrt{3})\,\ell_{\min}$, then $I_2<0$ as $\ell_1\to \ell_{\min}$, and $I_2$ becomes positive if $\ell_1$ is larger than some critical value smaller than ${1\over 2}(1+\sqrt{3})\,\ell_{\min}$.

We have seen that for a half interval on the sphere, $\tilde S[\mA]$ reproduces the entanglement entropy of \TTbar-deformed CFTs.  
If we extend this interpretation to the present case, 
the violation of subadditivity would suggest the possibility that the Hilbert space cannot be factorized as the product of local degrees of freedom, since $\tilde S[\mA_1]+\tilde S[\mA_2] < \tilde S[\mA_1\cup\mA_2]$ means that the union $\mA_1\cup\mA_2$ somehow contains more entanglement with its environment than the sum of the individual subsystems $\mA_i$. Similar violations of subadditivity have been observed in other interesting examples such as \cite{Li:2010dr,Kawamoto:2023nki}, and have been interpreted as a result of non-locality in the dual field theories. 
 
\item \emph{Strong subadditivity (SSA).}
Strong subadditivity is violated whenever the following function is negative
\eq{
I_3\equiv
\tilde S[{\mA_1 \cup \mA_2}] + \tilde S[{\mA_2 \cup \mA_3}] - \tilde S[{\mathcal A_1 \cup \mathcal A_2 \cup \mathcal A_3}] - \tilde S[{\mathcal A_2}]. \label{I3}
} 
Let us first consider three adjacent intervals $\mA_1$, $\mA_2$, and $\mA_3$ with unconstrained lengths $\ell_1$, $\ell_2$, and $\ell_3$. It is not difficult to find special cases where $I_3 < 0$. This occurs, for instance, when $\ell_1+\ell_2<\ell_{\min}$ and $\ell_1+\ell_2+\ell_3>\ell_{\min}$. On the other hand, when the lengths of the intervals are larger than $\ell_{\min}$, we can use the single interval expression $\tilde S_{\ell_i}$ for each of the four terms in \eqref{I3} such that
\eq{
I_3 = \tilde S_{\ell_1 + \ell_2}+\tilde S_{\ell_2 + \ell_3} - \tilde S_{\ell_1 + \ell_2 + \ell_3} - \tilde S_{\ell_2}.
}
It is straightforward to verify that all the partial derivatives $\p_{\ell_i}I_3$ are positive for $i=1,2,3$. We can also find a critical value $\tilde\ell$ of order $\ell_{\min}$ so that $I_3=0$ when the lengths of the intervals are equal to $\tilde\ell$. Then, strong subadditivity is satisfied as long as $\ell_i > \tilde\ell$ for all $i$, and it is violated whenever $\ell_i < \tilde\ell$ for all $i$.

In our discussions, the violation of strong subadditivity for $\mu > 0$ can be observed for intervals lying on a constant time slice. For $\mu < 0$, similar violations have been observed in \cite{Lewkowycz:2019xse}, but only when the intervals are boosted, such that they do not lie on the same time slice. 

\item {\it Infinitesimal version of SSA.}
Consider now three adjacent intervals $\mA_1, \mA_2$, and $\mA_3$ with lengths \eq{\ell_1=\ell_3={\epsilon \ell_x},\quad  \ell_2=\ell_x\, (1-\epsilon) >\ell_{\min}. }
Unlike the generic version of strong subadditivity, the infinitesimal one is guaranteed by concavity \eqref{eq:concavity} as $\epsilon \to 0$, namely  
\eq{ 
I_3= 2{\tilde S}_{\ell_x}-{\tilde S}_{\ell_x (1+\epsilon)}-{\tilde S}_{\ell_x (1-\epsilon)} \sim -\epsilon^2 \p^2_{\ell_x} \tilde S_{\ell_x} >0,\qquad
\epsilon\to 0.
}

\end{itemize}

These properties of the glue-on HRT formula are reminiscent of the entanglement entropy of an interval $\mA$ in a quantum field theory. This is not surprising, given that in the limit $\sqrt{\mu} \propto \epsilon \to 0$, \eqref{HEEpoincareAdS} reduces to the standard HRT formula of the AdS$_3$/CFT$_2$ correspondence. For finite values of $\mu$, we have seen that the glue-on HRT surface on a half interval on the sphere reproduces the entanglement entropy of \TTbar deformed CFTs with a finite UV cutoff determined by $\mu$. All of these results suggest a strong connection between the glue-on HRT formula and the entanglement entropy of \TTbar-deformed CFTs.


\section{HRT surfaces in glue-on AdS$_3$ on the cylinder}\label{s5}

In this section we construct glue-on HRT surfaces for spacelike intervals on the cylinder following the general prescription proposed in the previous section. In particular, we study HRT surfaces on both the vacuum and the nonrotating BTZ black hole. The fact that the spatial circle is compact in these cases leads to a novel interplay between the HRT surfaces and the minimum length of the interval. In particular, we will show that the topology of the glue-on BTZ background leads to a novel phase diagram for $\tilde S[\mA]$ as the size of the interval is changed. A general formula that is valid for small intervals on arbitrary stationary solutions of Einstein gravity is given in appendix \ref{se:general-cylinder}.


\subsection{Global AdS$_3$} \label{se:ERTAdS}

In this section we construct the HRT surface associated with an interval at a cutoff surface on a fixed-time slice of the glue-on version of global AdS$_3$. The signed area of this surface is expected to be related to the entanglement entropy of $T\bar T$-deformed CFTs with $\mu > 0$ on the vacuum. In particular, we will see that there is a minimum length of the interval below which the glue-on HRT surface ceases to exist, in agreement with the results of previous sections.

We begin by describing the glue-on version of global AdS$_3$. The metric can be written in the gauge introduced in section \ref{se:glueonAdS} as
\eq{
\quad ds^2 = \frac{\ell^2}{\zeta} \biggl( \frac{ d\zeta^2}{4\zeta(1+\zeta)}- (1+\zeta)\,dt^2  +d \vp^2 \biggr), \qquad  \vp \sim \vp + 2\pi,\qquad \zeta \ge \zeta_c, \label{auxglobalAdS}
}
where $\zeta_c < 0$ such that the cutoff surface is located in the AdS$_3^*$ region of the spacetime. Note that the signature of the spacetime changes at $\zeta = -1$ where the spacelike and timelike nature of the $\zeta$ and $t$ coordinates is exchanged. In analogy with the sphere foliation of AdS$_3$ considered in section \ref{se:HEEsphere}, we restrict the range of the radial coordinate in  \eqref{auxglobalAdS} to $\zeta \ge - 1$. Using the holographic dictionary  \eqref{dictionary}, we see that this range of $\zeta$ reproduces the bound on the deformation parameter of \TTbar-deformed CFTs on a cylinder of radius $R = \ell$ \eqref{criticalmu}.

\begin{figure}[ht!]
{\centering
\subfloat[$\ell_\mA > \ell_{\min}$ \label{fig:globalRTa}]{
 \includegraphics[scale=0.478]{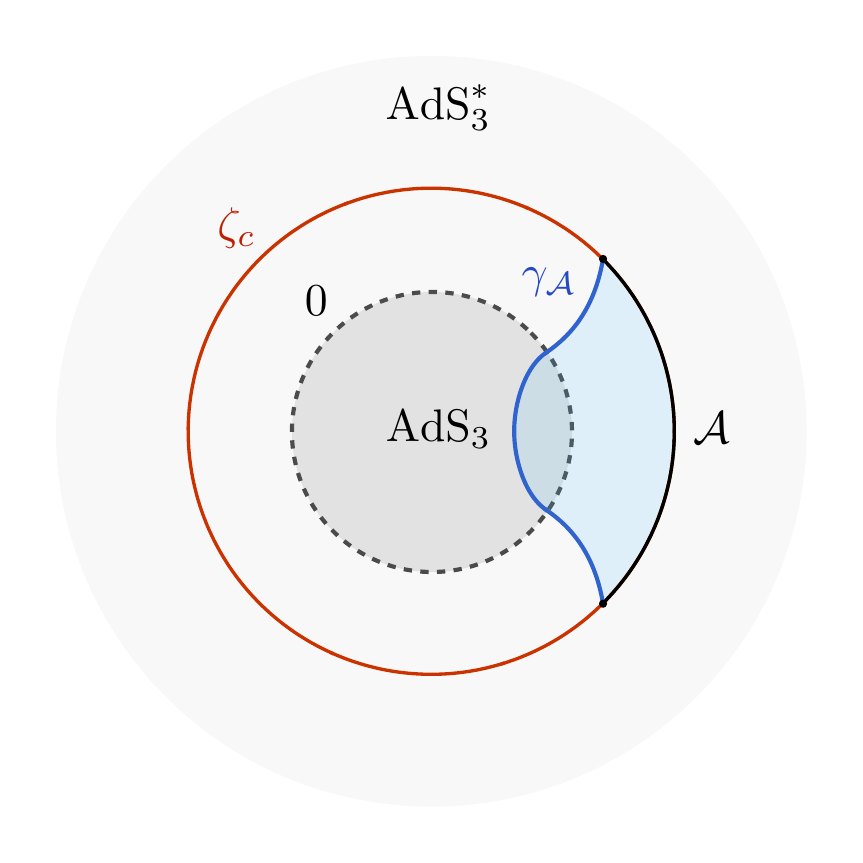}
}\hfill
\subfloat[$\ell_\mA = \ell_{\min}$\label{fig:globalRTb}]{
 \includegraphics[scale=0.478]{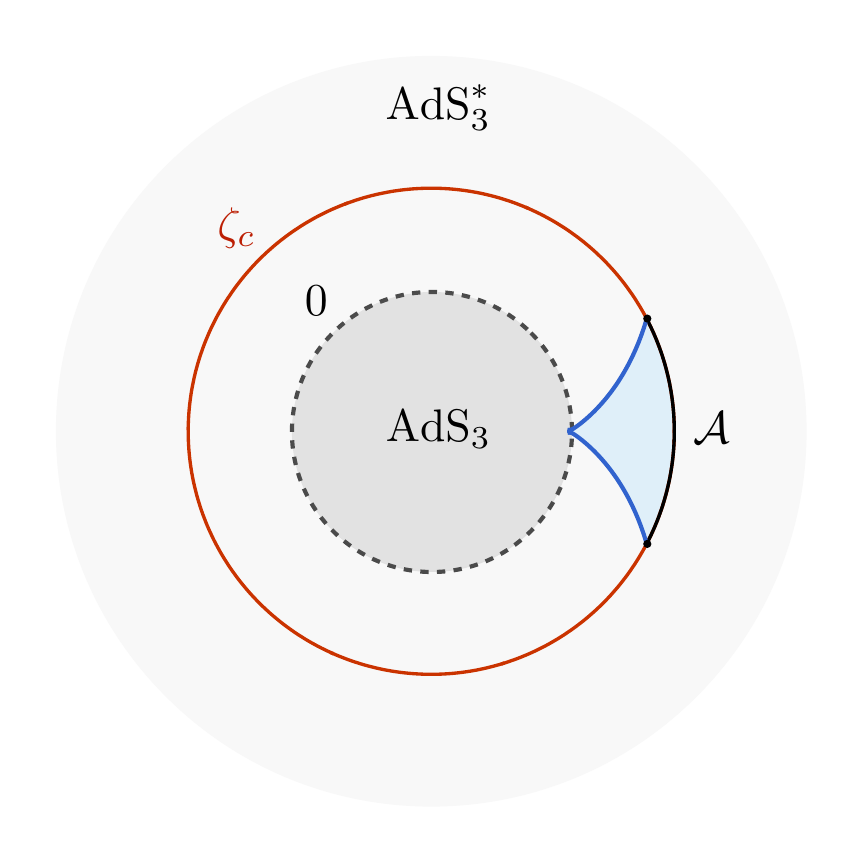}
}
}
\caption{Fixed-time slices of the glue-on version of global AdS$_3$ and the HRT surfaces $\gamma_\mA$ (blue) associated with intervals $\mA$ at the cutoff surface (red) in the AdS$_3^*$ region. When the length of the interval $\ell_\mA$ equals $\ell_{\min}$, the AdS$_3$ part of the candidate HRT surface shrinks to a point and the AdS$_3^*$ part becomes lightlike.}
\label{fig:globalRT}
\end{figure}

Let us now consider an interval $\mathcal A$ on the cutoff surface $\zeta=\zeta_c$. For convenience, we assume that the interval lies on a fixed-time slice so that it can be parametrized by its angular coordinate by
\eq{
\mathcal A =\big \{ \varphi \, |\, \varphi \in [-\varphi_\mA, \varphi_\mA] \big\}.\label{globalAdSinterval}
}
The total length of the interval is then $\ell_\mA =  2 \ell \varphi_\mA$. We are interested in finding the extremal spacelike surface $\gamma_\mA$ with the smallest signed area that is anchored to the interval \eqref{globalAdSinterval} at the cutoff surface. Following the discussion in the previous section, we can construct the glue-on HRT surface in the following way. We first consider a standard HRT surface in the AdS$_3$ region of the glue-on AdS$_3$ spacetime. In the present case, it is more convenient to parametrize the HRT surface in terms of the angular, instead of the radial, coordinate. The function describing the HRT surface is then single valued. In order to obtain the glue-on HRT surface we continue the value of the angular coordinate such that the HRT surface crosses to the AdS$_3^*$ region of the spacetime. The glue-on HRT surface can be shown to be given by (see fig.~\ref{fig:globalRTa} for an illustration)
\eq{
\gamma_{\mA}: \quad \zeta(\vp)=(1+\zeta_c)\,\frac{\cos^2\vp}{\cos^2\vp_\mA} - 1. \label{glueonRTglobalAdS}
}

The glue-on HRT surface \eqref{glueonRTglobalAdS} consists of two segments in AdS$_3^*$ and an HRT surface in AdS$_3$. The latter is anchored to the following auxiliary interval at the asymptotic boundary
\eq{
\hat {\mathcal A} = \big\{ \varphi \, |\, \varphi \in \bigl [- \varphi_{\hat\mA}, \varphi_{\hat\mA}\, \big] \big\}, \qquad (1+\zeta_c)\cos^2 \vp_{\hat\mA} = \cos^2 \vp_\mA. \label{hatintervalglobalads}
}
Decreasing the size of the interval $\mA$ at the cutoff surface, decreases the size of the interval $\hat{\mA}$ at the asymptotic boundary. In particular, the HRT surface \eqref{glueonRTglobalAdS} does not exist when the size of the interval $\hat{\mA}$ shrinks to zero. At this point, the AdS$_3$ part of $\gamma_\mA$ disappears, and the two segments in AdS$_3^*$ become lightlike. As a result, we find that in this case there is also a minimum length of the interval for which the glue-on HRT surface is guaranteed to exist that is given by
\eq{
\ell_\mA > \ell_{\min}, \qquad  \ell_{\min}  = 2\ell \arcsin  \big( \sqrt{-\zeta_c}\, \big) = 2\ell \arcsin \bigg( \sqrt{ \frac{c \mu}{3 \ell^2} }\bigg). \label{minimumlengthglobal}
}
In particular, note that for small values of $\mu$ we have $\ell_{\min}^2 = 4c\mu/3 + \mathcal O (\mu^2)$, which reduces to the result obtained for Poincar\'e AdS$_3$ in section \ref{se:ERTPoincare}. On the other hand, when $\mu < 0$, the spacelike geodesic $\gamma_\mA$ lies on the AdS$_3$ part of the spacetime and is described by \eqref{glueonRTglobalAdS} with $\vp < \vp_{\hat\mA}$. In this case, $\gamma_\mA$ corresponds to an HRT surface attached to the cutoff surface $\zeta=\zeta_c>0$ and there is no minimum length of the interval.

The signed area of the glue-on HRT surface \eqref{glueonRTglobalAdS} can be written as
\eq{
\widetilde{\mathrm{Area}}[\gamma_\mA] &= 2  \int_0^{\vp_{\hat\mA} -  \epsilon} \sqrt{g_{\mu\nu} \partial_\vp x^{\mu} \partial_\vp x^{\nu}} \,d\vp    -   2 \int_{\vp_{\hat\mA} + \epsilon}^{\vp_\mA} \sqrt{g_{\mu\nu} \partial_\vp x^{\mu} \partial_\vp x^{\nu}} \,d\vp  \label{glueonRTglobalAdSint1} \\
 &= 2\ell \bigg( \int_0^{\vp_{\hat\mA} -  \epsilon}  +  \int_{\vp_{\hat\mA} +  \epsilon}^{\vp_\mA}  \bigg) \bigg(\frac{ \tan \vp_{\hat\mA}\,d\vp}{\cos^2 \vp \sec^2 \vp_{\hat\mA} - 1} \bigg), \label{glueonRTglobalAdSint2}
}
where $\epsilon \to 0^+$ regulates the divergence of the integral near the asymptotic boundary at $\vp = \vp_{\hat\mA}$. The first term in \eqref{glueonRTglobalAdSint1} is the area of the HRT surface attached to the asymptotic boundary of AdS$_3$, while the second term corresponds to the area of the AdS$_3^*$ part of \eqref{glueonRTglobalAdS}. Evaluating the integral, we find that the area of $\gamma_\mA$ is finite, independent of the cutoff $\epsilon$, and given by
\eq{
\widetilde{\mathrm{Area}}[\gamma_\mA] & = \ell\log\bigg( \frac{\thinmspace\sin\thinmspace(\vp_\mA + \vp_{\hat\mA})}{\sin|\vp_\mA - \vp_{\hat\mA}|}\bigg).
}
Using the holographic dictionary \eqref{dictionary}, together with the relationship between $\vp_\mA$ and $\vp_{\hat\mA}$ in \eqref{hatintervalglobalads}, the glue-on HRT formula yields
\eq{
\tilde S[\mA] = \frac{\widetilde{\textrm{Area}}[\gamma_\mA]}{4G} =	\left\lbrace
\begin{aligned}
\,
&\frac{c}{3} \arccosh\biggl(\sqrt{\frac{3 \ell^2}{c \mu}} \sin \frac{\ell_\mA}{2\ell} \biggr), &&\mu > 0,
	\\[.5ex]
&\frac{c}{3} \arcsinh\biggl(\sqrt{-\frac{3 \ell^2}{c \mu}} \sin  \frac{\ell_\mA}{2\ell} \biggr),\quad &&\mu < 0, 
\end{aligned}
\right. \label{HEEglobalAdS}
}
where we have also included the result for $\mu < 0$ for completeness. 

Let us now comment on a few features of \eqref{HEEglobalAdS}. First, when $\mu > 0$, the right hand side of \eqref{HEEglobalAdS} is valid only for $\ell_\mA > \ell_{\min}$. As discussed earlier, the glue-on HRT surface does not exist for $\ell_\mA \le \ell_{\min}$, in which case we define $\tilde S[\mA]=0$. Altogether, the single interval result is still non-negative, which implies that $\tilde S$ is non-negative for all intervals in global AdS$_3$. This is the same behavior observed for Poincar\'e AdS$_3$ in section \ref{se:ERTPoincare}. Since the spatial circle is contractible in the AdS$_3$ region of the spacetime, an interval with size $\ell_\mA > \ell_{\min}$ shares the same HRT surface of its complement with size $\ell_{\mA^c} = 2\pi \ell-\ell_{\min}$. As a result, we have $\tilde S[\mA]=\tilde S[\mA^c]$, which is analogous to the purity condition described in section \ref{se4:featuresHRT}. The existence of a minimum length then implies the existence of a maximal length $\ell_{\max} = 2\pi \ell-\ell_{\min}$ beyond which the glue-on HRT surface ceases to exist. Finally, note that when the deformation parameter saturates the bound \eqref{criticalmu}, i.e.~when it takes the critical value $\mu_c= 3\ell^2/c$, the minimum length of the interval \eqref{minimumlengthglobal} becomes half of the size of the system, namely
\eq{
\ell_{\min} \big|_{\mu = \mu_c} = \pi \ell=\ell_{\max}.
}
Consequently, there are no glue-on HRT surfaces for any interval when $\mu$ reaches its critical value. These features of the glue-on HRT formula on global AdS$_3$ are reminiscent of the behavior of the entanglement entropy on a pure state and it would be interesting to explore further the relationship between \eqref{HEEglobalAdS} and the vacuum entanglement entropy of \TTbar-deformed CFTs.


\subsection{BTZ black holes}
\label{se:ERTBTZ}

In this section we study the consequences of adding temperature to the glue-on HRT formula. In this case, the HRT formula is given by the signed area of an extremal surface attached to the endpoints of an interval on a cutoff surface in the glue-on version of the BTZ black hole. We will show that, in analogy with the holographic entanglement entropy of two-dimensional CFTs, $\tilde{S}[\mA]$ of an interval $\mA$ differs from that of its complement $\mA^c$, a result that is related to the fact that BTZ is not a pure state. In particular, we will show that the signed area can lead to situations where the extremal surface homologous to $\mA$ always dominates as we increase the size of the interval.

Let us consider the glue-on version of the nonrotating BTZ black hole. The metric can be written in the coordinates used in \eqref{se2:glueonAdS} as
\eq{
ds^2 = \frac{\ell^2}{\zeta} \Bigl[ \frac{ d\zeta^2}{4\zeta(1-r_+^2\zeta)}- (1-r_+^2\zeta)\,dt^2  +d \vp^2 \Big], \qquad  \vp \sim \vp + 2\pi, \quad \zeta \ge \zeta_c,
}
where $\zeta = \zeta_c \ge -1$ is the location of the cutoff surface, the horizon is located at $\zeta^{-1} = r_+^2$, and $r_+ = 2\pi\ell/\beta_t$ where $\beta_t$ is the inverse temperature of the black hole. Due to the $(1-r_+^2\zeta)$ factor in the $g_{tt}$ component of the metric, the inverse temperature $\beta$ for the \TTbar-deformed CFT living at the cutoff surface is related to $\beta_t$ via \cite{Apolo:2023vnm}
\eq{
\beta = \beta_t \sqrt{1 - r_+^2\zeta_c} 
}
The location of the cutoff surface $\zeta_c$ is related to the \TTbar deformation parameter via the holographic dictionary \eqref{dictionary}. 

\begin{figure}[ht!]
{\centering
\subfloat[\label{fig:BTZRTa}Connected]{
 \includegraphics[scale=0.478]{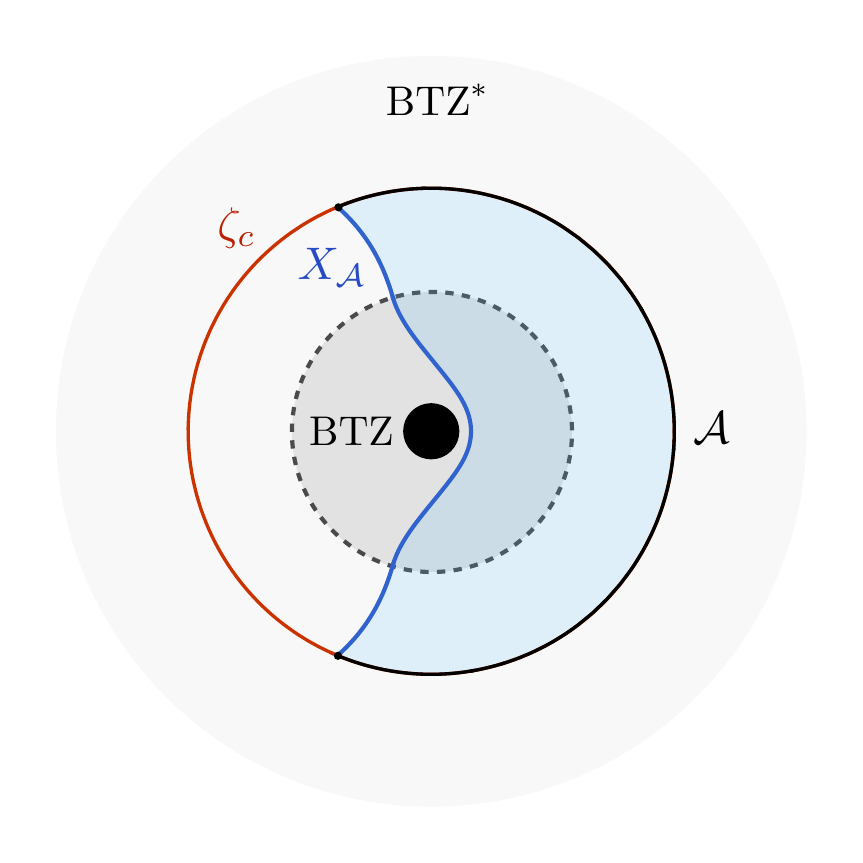}
}\hfill
\subfloat[\label{fig:BTZRTb}Disconnected]{
 \includegraphics[scale=0.478]{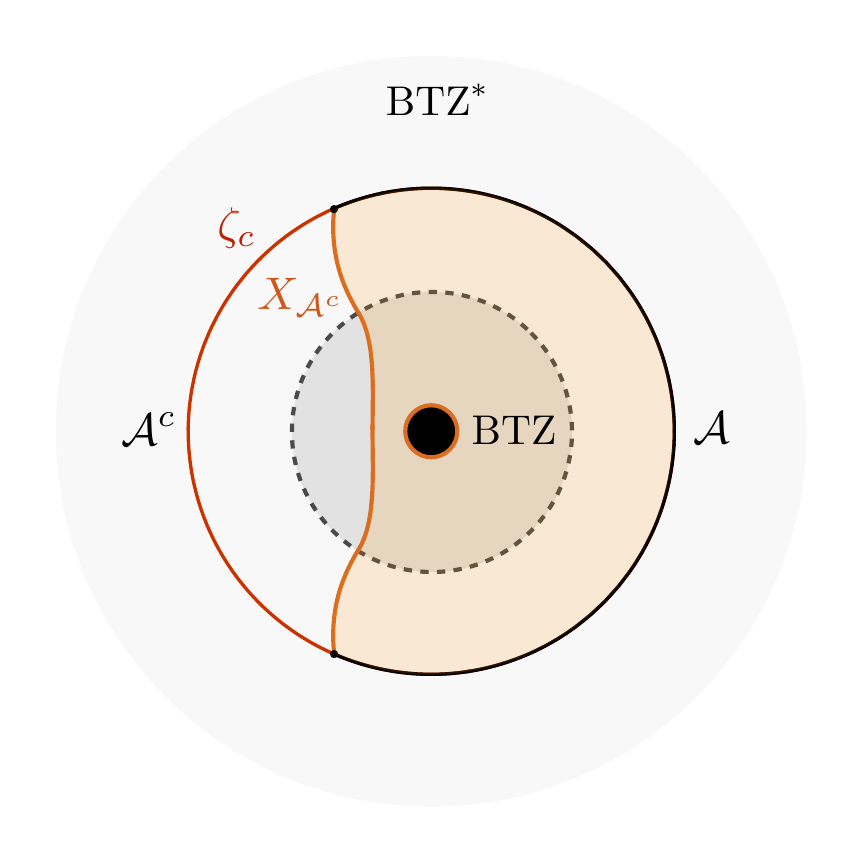}
}
}
\caption{Extremal surfaces associated with an interval $\mA$ at the cutoff surface (red) on a fixed-time slice of a glue-on BTZ spacetime. The connected surface (a) is homologous to the interval $\mA$, while the disconnected one (b) consists of the union of the circle around the black hole horizon (black disk) and the surface homologous to the complement $\mA^c$. 
}
\label{fig:BTZRT}
\end{figure}

We now consider an interval $\mA$ on the cutoff surface $\zeta = \zeta_c$ on the BTZ$^*$ ($\zeta < 0$) region of the spacetime at a fixed-time slice. The interval is parametrized by the angular coordinate as in \eqref{globalAdSinterval} and the extremal surface $X_\mA$ homologous to $\mA$ is given by
\eq{
X_{\mA}\colon\quad  \zeta(\vp) = \bigg(\zeta_c - \frac{1}{r_+^2}\bigg) \bigg(\frac{\cosh^2r_+\vp}{\cosh^2r_+\vp_\mA}+1 \bigg).\label{ERTBTZ}
}
The extremal surface $X_\mA$ consists of two parts lying in the BTZ and BTZ$^*$ regions of the glue-on spacetime. As illustrated in fig.~\ref{fig:BTZRTa}, the BTZ part of $X_\mA$ consists of a standard HRT surface that lies outside of the horizon of the BTZ black hole and attaches to an auxiliary interval $\hat\mA$ at the asymptotic boundary. The interval $\hat\mA$ can be parametrized by
\eq{
\hat {\mathcal A} = \big\{ \varphi \, |\, \varphi \in \bigl[ - \varphi_{\hat\mA}, \varphi_{\hat\mA\,} \big] \big\}, \qquad (1-r_+^2\zeta_c)\cosh^2 (r_+ \vp_{\hat\mA}) =  \cosh^2 (r_+ \vp_\mA). \label{hatintervalBTZ}
}
As a result, there is a minimum length of the interval $\mA$ that is necessary for the existence of the glue-on HRT surface such that
\eq{
\ell_\mA > \ell_{\min}, \qquad \ell_{\min}  =  \frac{2\ell}{r_+}\arcsinh \bigg(r_+ \sqrt{\frac{c \mu}{3 \ell^2}}\, \bigg) \le 2\ell, \label{minimumlengthBTZ}
}
where the last inequality follows from the bound on the deformation parameter \eqref{criticalmu}. Notably, in this case the minimum length depends on the size of the black hole, decreasing as $r_+$ is increased.
In particular, from \eqref{minimumlengthBTZ} we learn that $\ell_{\min}< 2\pi\ell-\ell_{\min}$. It is not difficult to verify that the extremal surface \eqref{ERTBTZ}, the interval \eqref{hatintervalBTZ}, and the minimum length \eqref{minimumlengthBTZ} reduce to the corresponding quantities on global AdS$_3$ after the analytic continuation $r_+ = i$ that turns the nonrotating BTZ black hole into the global AdS$_3$ spacetime. 

The signed area of the extremal surface \eqref{ERTBTZ} can be written as 
\eq{
\sArea[X_\mA] &= 2  \int_0^{\vp_{\hat\mA} - \epsilon} \sqrt{g_{\mu\nu} \partial_\vp x^{\mu} \partial_\vp x^{\nu}} \,d\vp    -   2 \int_{\vp_{\hat\mA} +  \epsilon}^{\vp_\mA} \sqrt{g_{\mu\nu} \partial_\vp x^{\mu} \partial_\vp x^{\nu}} \,d\vp  \notag  \\
 &= 2\ell \bigg( \int_0^{\vp_{\hat\mA} -  \epsilon}  +  \int_{\vp_{\hat\mA} + \epsilon}^{\vp_\mA}  \bigg) \bigg(\frac{ r_+ \tanh( r_+ \vp_{\hat\mA})\,d\vp}{1 - \cosh^2 (r_+ \vp) \mop{sech}^2 (r_+ \vp_{\hat\mA})} \bigg) \notag \\
& = \ell\log\bigg( \frac{\sinh  ( r_+ (\vp_\mA + \vp_{\hat\mA}))}{\sinh ( r_+ |\vp_\mA - \vp_{\hat\mA}|)}\bigg),\label{sarea}
}
where $\epsilon \to 0^+$ is the UV cutoff and the absolute value in the last line guarantees that the expression is valid for cutoff surfaces that lie either in the BTZ or the BTZ$^*$ regions of the spacetime. 

Due to the existence of a black hole horizon, the homologous condition for an interval ${\mA}$ differs from that of its complement ${\mA^c}$, and we expect $\gamma_{\mA}$ to be different from $\gamma_{\mA^c}$. This is related to the thermal nature of the dual state on the field theory side, and can be seen directly from \eqref{sarea}, as the expression is not invariant when $\vp_\mA$ is exchanged by its complement $\pi - \vp_\mA$. If the interval is small enough such that $\ell_\mA < \ell_{\min}$, then there is no extremal surface homologous to $\mA$, and hence $\tilde S[\mA]=0$. On the other hand, when $\ell_\mA > \ell_{\min}$, there are two extremal surfaces homologous to $\mA$ as illustrated in fig.~\ref{fig:BTZRT}.  One surface, $X_\mA$, is connected and homotopic to $\mA$. The other surface is disconnected and consists of the union of the BTZ horizon and $X_{\mA^c}$, the latter of which is homotopic to the complement $\mA^c$. As $\ell_\mA$ approaches $2\pi\ell-\ell_{\min}$ from below, the extremal surface $X_{\mA^c}$ shrinks towards the BTZ$^*$ region of the spacetime and ceases to exist once this value is reached. As a result, in the case $\ell_\mA \ge 2\pi\ell-\ell_{\min}$, the disconnected surface consists only of the BTZ horizon.

Using \eqref{sarea} and \eqref{hatintervalBTZ}, the contribution of \eqref{ERTBTZ} to the glue-on HRT formula for a generic value of $\ell_\mA > \ell_{\min}$ is given by
\eq{
\tilde S_{\ell_\mA} \equiv  \frac{c}{3} \arccosh\bigg( \frac{1}{r_+} \sqrt{\frac{3 \ell^2 }{c \mu}} \sinh  \frac{r_+ \ell_\mA}{2\ell}  \bigg) =  \frac{c}{3} \arccosh\biggl(  \sinh  \frac{r_+ \ell_\mA}{2\ell} \Big/\sinh  \frac{r_+ \ell_{\min}}{2\ell} \bigg).
}
In terms of this quantity, the contribution of the connected and disconnected surfaces to the glue-on HRT formula read
\eq{
\text{connected:}\quad \til S_{\text{con}} &= \left\lbrace
\begin{aligned}
\,&0, && \ell_\mA<\ell_{\min}, \\
& \tilde S_{\ell_\mA}, \qquad\qquad\qquad\,\,\,  && 
\ell_\mA \ge \ell_{\min}, 
\end{aligned}
\right. \label{HEEBTZXA}
\\[1ex]
\text{disconnected:}\quad \til S_{\text{dis}} &= \left\lbrace
\begin{aligned}
\ &\tfrac{c}{3} \pi r_+ +  \tilde S_{2\pi \ell - \ell_\mA} ,  \quad && \ell_\mA<2\pi \ell -\ell_{\min} ,\\
&\tfrac{c}{3} \pi r_+,
&&\ell_\mA \ge 2\pi \ell -\ell_{\min},
\end{aligned}\label{HEEBTZXAc}
\right.
}
where $\tfrac{c}{3}\pi r_+$ in \eqref{HEEBTZXAc} is the entropy of the black hole. 

\begin{figure}[ht!]
\hspace{-1.5em}
\centering%
\mbox{
\parbox[c]{.33\textwidth}{%
\subfloat[
	$r_1 < r_+$%
	\label{fig:BTZ-one-phase}
]{%
	\includegraphics[scale=1]{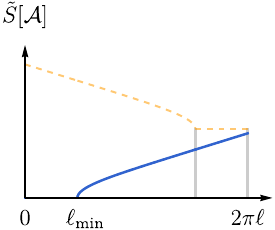}
}
}
\parbox[c]{.33\textwidth}{%
\subfloat[
	$r_2 < r_+ < r_1$%
	\label{fig:BTZ-two-phases}
]{%
	\includegraphics[scale=1]{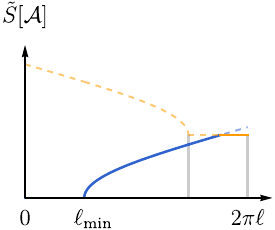}
}
}
\parbox[c]{.33\textwidth}{%
\subfloat[
	$r_+ < r_2$%
	\label{fig:BTZ-three-phases}
]{\parbox[c]{.33\textwidth}{
    \includegraphics[scale=1]{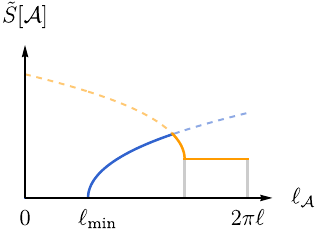}
}}
}
}
\caption{%
The value of $\tilde S[\mA]$ for the dominant glue-on HRT surface as a function of the interval length $\ell_\mA$. The blue line corresponds to the contribution of the connected extremal surface while the orange line is the contribution of the disconnected one. As shown in (a), for a large enough horizon size $r_+$, the disconnected surface never dominates, and $\tilde S[\mA]$ is always smaller than the black hole entropy.
}
\label{fig:BTZ-phases}
\end{figure}

The value of $\tilde S[\mA]$ is given by the extremal surface with the minimum signed area. The connected contribution $\tilde S_{\text{con}}$ is zero until $\ell_\mA$ reaches $\ell_{\min}$ and monotonically increases thereafter (see fig.~\ref{fig:BTZ-one-phase}). On the other hand, the disconnected contribution $\tilde S_{\text{dis}}$ monotonically decreases until $\ell_\mA$ reaches $2\pi\ell-\ell_{\min}$, where $\tilde S_{\text{dis}}$ is given by the black hole entropy. Interestingly, it is possible for the maximum value of $\tilde S_{\text{con}}$ (reached when $\ell_\mA=2\pi\ell$), to be less than the black hole entropy, case in which there is no phase transition and the HRT surface is always connected.  The condition for the appearance of a phase transition is 
\eq{
r_+^{2}  \coth^2 {\pi r_+ } \le \frac{3\ell^2}{c\mu}. \label{nophtr}
}
For a fixed value of $\mu$, this inequality is saturated at a critical value of $r_+$ that we denote by $r_1$, and it is satisfied for all $r_+ < r_1$. Consequently, when $r_+ > r_1$ there is no phase transition and $\tilde S[\mA] = \til S_{\text{con}}$ for all intervals. 

On the other hand, when $r_+ < r_1$, a phase transition necessarily occurs. In this case, the phase transition may occur when the size of the interval $\ell_\mA$ is smaller or greater than $2\pi\ell-\ell_{\min}$. When $\ell_\mA < 2\pi\ell-\ell_{\min}$, a phase transition is possible if the maximum contribution from the connected extremal surface in this range is greater than the black hole entropy. This is equivalent to the condition
\eq{
r_+^2 (4\coth^2 \pi r_+ - 1) \le \frac{3\ell^2}{c\mu}. \label{trans2}
}
This inequality is saturated at a value of $r_+$ that we denote by $r_2$, and is satisfied for all $r_+ < r_2$. Since the left hand side of \eqref{trans2} is greater than that of \eqref{nophtr}, we see that the critical values satisfy $r_1 > r_2$. This is consistent with the appearance of a phase transition in this regime.

We conclude that the glue-on HRT formula depends on the size of the black hole, i.e.~on the temperature of the thermal state, and is always positive semi-definite. As illustrated in fig.~\ref{fig:BTZ-phases}, there are three types of phase diagrams and the glue-on HRT formula features the following transitions
\begin{itemize}
\item When $r_1 < r_+$, where $r_1$ saturates \eqref{nophtr}, the value of $\tilde S[\mA]$ is given by \eqref{HEEBTZXA} for all values of $\ell_\mA$ and we find
\eq{
\tilde S[\mA] = \tilde S_{\text{con}}. 
}
\item When $r_2 < r_+ < r_1$, where $r_2$ saturates the bound \eqref{trans2}, a phase transition occurs at a critical length $\tilde{\ell} > 2\pi \ell-\ell_{\min}$, such that
\eq{
\tilde S[\mA]=\left\lbrace
\begin{aligned}&\tilde S_{\text{con}}, \qquad&&\ell_\mA\le\tilde \ell\\
&\frac{c}{3}\,\pi r_+, && \ell_\mA\ge \tilde \ell \ge 2\pi\ell -\ell_{\min},
\end{aligned}
\right.
}
where $\tilde\ell$ is determined by equating the first and second lines.

\item When $r_+ < r_2$, a phase transition occurs at a critical length $\ell_{\min}<\tilde{\ell}<2\pi \ell-\ell_{\min}$ and we have 
\eq{
\tilde S[\mA]=\left\lbrace
\begin{aligned}&\tilde S_{\text{con}}, \qquad &&\ell_\mA\le \tilde \ell\le 2\pi\ell -\ell_{\min},\\
&\tilde S_{\text{dis}}, && \ell_\mA\ge \tilde \ell,
\end{aligned}
\right.
}
where $\tilde S_{\text{dis}}$ is given in \eqref{HEEBTZXAc}, which is greater than the black hole entropy for $\ell_\mA < 2\pi\ell -\ell_{\min}$, as shown in fig.~\ref{fig:BTZ-three-phases}.

\end{itemize}

Finally, we note that when $\mu < 0$, the glue-on HRT surface reduces to a standard HRT surface with a finite cutoff. In this case there is no minimum length of the interval and the phase diagram is similar to that of CFT$_2$ in a thermal state. In this case, we have
\eq{
\tilde S[\mA]   =	\left\lbrace
\begin{aligned}
&\frac{c}{3} \arcsinh\bigg(\frac{1}{r_+} \sqrt{-\frac{3\ell^2 }{c \mu}} \sinh  \frac{r_+ \ell_\mA}{2\ell}  \bigg),  \qquad  \qquad\qquad\quad\quad\, \ell_\mA < \til \ell,
	\\[.5ex]
& \frac{c}{3} \pi r_+ + \frac{c}{3} \arcsinh\bigg( \frac{1}{r_+} \sqrt{-\frac{3\ell^2 }{c \mu}} \sinh \frac{r_+ (2\pi\ell - \ell_\mA)}{2\ell}  \bigg), \quad \ell_\mA \ge \til \ell,
\end{aligned}
 \right. \qquad  \mu < 0,  \notag
}
where $\tilde \ell$ is the length of the interval for which the two lines are equal to each other.

\bigskip

\section*{Acknowledgments}
We are grateful to Bin Chen, Bartek Czech, Kanato Goto, Xia Gu, Monica Guica, Kangning Liu, Reiko~Liu, Dominik Neuenfeld, Cheng Peng, Xiao-Liang Qi, Andrew Rolph, Shan-Ming Ruan, Tadashi Takayanagi, Huajia Wang, Jie-Qiang~Wu and Yuan Zhong for helpful discussions. 
The work of LA was supported in part by the Dutch Research Council (NWO) through the Scanning New Horizons programme (16SNH02). The work of PXH, WXL, and WS is supported by the national key research and development program of China No.~2020YFA0713000. LA thanks the Asia Pacific Center for Theoretical Physics (APCTP) for hospitality during the focus program ``Integrability, Duality and Related Topics", as well as the Korea Institute for Advanced Study (KIAS) for hospitality during the ``East Asia Joint Workshop on Fields and Strings 2022'', where part of this work was done. WXL and WS thank the Yukawa Institute for Theoretical Physics (YITP) for hospitality during the ``YIPQS long-term workshop on Quantum Information, Quantum Matter and Quantum Gravity'' (YITP-T-23-01), where part of this work was completed.


\appendix

\section{Renormalized entropy from endpoints} \label{se:renormailzed-EE}

In this appendix we revisit the derivation of the holographic entanglement entropy of a half interval in  \TTbar-deformed CFTs on the sphere.

As described in section \ref{se:sphere-field-theory}, the sphere partition function of \TTbar-deformed CFTs $Z_\mu(a)$ depends on an integration constant $a$ with the dimension of length that is related to the renormalization scale of the theory. The choice $a = \sqrt{c|\mu| / 3}$ leads to an exact match between the field theory calculation \eqref{EEsphere} and the holographic result \eqref{HEEEAdS}, and furthermore reproduces the result of \cite{Donnelly:2018bef} when $\mu < 0$. The relationship between the length scale $a$ and the deformation parameter $\mu$ is a natural one from the point of view of cutoff/glue-on AdS holography and the UV/IR relation of the AdS/CFT correspondence. This follows from the fact that changing the location of the asymptotic boundary of the bulk spacetime, which is determined by $\mu$, is interpreted as changing the UV cutoff $a$ of the dual field theory, such that there is a single scale specified by $\mu$. 

On the other hand, it would also be interesting to study the \TTbar deformation with an independent UV length scale, which is natural from the field theory perspective. In this case the integration constant $a$ is decoupled from the deformation parameter $\mu$. Using the general formula \eqref{Zsol} for the partition function $Z_\mu (a)$, we find that for generic $a$, the entanglement entropy is given by \cite{Li:2020zjb}
\eq{
S_a[\mA]
= \frac{c}{3} \log \bigg[\frac{L}{a}   \bigg(1+\sqrt{1-\frac{c \mu }{3  L^2}}\, \bigg) \bigg].
\label{eq:renormalized-entropy}
}
We see that the integration constant $a$ now enters the entanglement entropy. This can be understood as a renormalized quantity, where the UV cutoff is always tuned to the length scale $a$. Unlike the previous result \eqref{EEsphere}, the entropy $S_a[\mA]$ is analytic in $\mu$, and it admits a direct $\mu \to 0$ limit, where  $S_a[\mA] \to ({c}/{3}) \log\,({2L}/{a})$ and $a$ is simply identified with the UV cutoff of the original CFT.

It would then be interesting to identify the holographic prescription for the renormalized entropy \eqref{eq:renormalized-entropy}. In order to incorporate an independent boundary radius $L$, we replace $\zeta \to (\ell^2/L^2)\,\zeta$ in the background \eqref{glueSphere}. Under this rescaling, the metric is given by:
\eq{
\quad ds^2 =  \frac{\ell^2  d \zeta^2}{4\zeta^2(1 + \frac{\ell^2}{L^2}\zeta)} + \frac{L^2}{\zeta} \big( d\theta^2 + \sin\theta^2\,d\phi^2\big), \qquad \zeta \ge \zeta_c=- \frac{c \mu}{3 a^2}.
\label{eq:sphere-rescaled}
}
The two scales $L$ and $\ell$ are decoupled here, unlike in previous sections. Consider the partition function $Z^{(n)}$ of the $n$-cover of the sphere, which is smooth in the bulk but has conical singularities with angle $2\pi n$ at the endpoints. The holographic entanglement entropy is given by the bulk extension of the replica trick \eqref{eq:replica-trick}. As shown in \cite{Lewkowycz:2013nqa}, the holographic entanglement entropy of  a QFT is given by
\eq{
	 S[\mA]
	= \bigg(1 - n \frac{d}{dn} \bigg) \log Z^{(n)}_\mu \big|_{n = 1}
	= \pdd{n} I^{(n)}\big|_{n = 1},
\label{eq:bulk-replica-trick}
}
where $I^{(n)}$ is the gravitational action of a regularized conical singularity around the $\mbb{Z}_n$ fixed loci. 

Let us assume that the glue-on HRT formula for the holographic entanglement entropy of \TTbar-deformed CFTs can be obtained as in \eqref{eq:bulk-replica-trick} such that
\eq{
	\tilde S[\mA]
	= \bigg(1 - n \frac{d}{dn} \bigg) \log Z^{(n)}_\mu \big|_{n = 1}
	= \pdd{n} I^{(n)}\big|_{n = 1}.
\label{eq:bulk-replica-trick2}
}
As explained in \cite{Apolo:2023vnm}, the renormalized action can include a boundary Weyl counterterm
\eq{
I_{W}(\zeta_c) = -\frac{\ell}{32\pi G}
\log |\zeta_c|
\int_{\mathcal N_c} d^2x \sqrt{\gamma} \,  \mathcal{R}[\gamma], \label{eq:weyl-counterterm}
}
where $\mathcal{R}[\gamma]$ is the Ricci scalar of the cutoff surface $\mathcal N_c$ computed with respect to the boundary metric $\gamma_{ij}$. Here, we focus on the correction to $\tilde S[\mA]$ originating from the Weyl counterterm. Since $I_{W}(\zeta_c)$ is a boundary term, it depends on the location of the cutoff surface, and its contribution is concentrated within small disks $D_i$ around the $\mbb{Z}_n$ fixed points at $\mathcal N_c$, i.e.~around the endpoints of the interval $\mA$. We then find \cite{Fursaev:1995ef,Lewkowycz:2013nqa}
\eq{
	 I^{(n)}_{W}(\zeta_c)
	&= -\frac{\ell}{32\pi G}
        \log |\zeta_c|
		\int_{\cup_i D_i} d^2x \sqrt{\gamma} \,
		\mathcal{R}[\gamma] \notag \\[1ex]
	&= -\frac{\ell}{32\pi G}
        \log |\zeta_c|
		\, 4\pi(1-n) \times 2,
}
where $4\pi(1-n)$ is the contribution from one conical singularity, and we have two of them at each endpoint of $\mA$. The correction to the entanglement entropy originating from the addition of the Weyl counterterm is thus given by
\eq{
	\tilde S_W [\zeta_c] \equiv \pd_n I_W^{(n)}(\zeta_c)\big|_{n = 1}
	= \frac{\ell}{4G}
		\log \abs{\zeta_c}
	= \frac{c}{3}
		\log \sqrt{\frac{c\mspace{1mu}|\mu|}{3\ell^2}}. \label{eq:entropy-corrections}
}
This is precisely the difference between the entropy \eqref{HEEEAdS} and the renormalized entropy \eqref{eq:renormalized-entropy} provided that we identify $a = \ell$ such that
\eq{ 
 \tilde S[\mA] + \tilde S_W [\zeta_c]  = 
	&= \frac{c}{3} \log \bigg[\frac{L}{\ell}   \bigg(1+\sqrt{1-\frac{c \mu }{3 L^2}}\, \bigg) \bigg]
	= S_\ell[\mA].
} 

Finally, note that in \eqref{eq:weyl-counterterm} we have chosen the standard normalization for the Weyl counterterm. However, as emphasized in \cite{Anastasiou:2020zwc}, there is actually an ambiguity in the coefficient of $I_W(\zeta_c)$, which corresponds to the renormalization scheme dependence of the theory. In particular, it is equally valid to consider
\eq{
I_W\big(\tfrac{\ell^2}{a^2}\zeta_c\big) = -\frac{\ell}{32\pi G} \log \Big|\frac{\ell^2}{a^2}\zeta_c\Big| \int_{\mathcal N_c} d^2x \sqrt{\gamma} \,  \mathcal{R}[\gamma].
}
This leads to the following correction to the entanglement entropy
\eq{
	\tilde S_W \big[ \tfrac{\ell^2}{a^2}\zeta_c\big] = \pd_n I_W^{(n)}\big(\tfrac{\ell^2}{a^2}\zeta_c \big)\big|_{n = 1}
	= \frac{\ell}{4G}
		\log \Big|\frac{\ell^2}{a^2}\zeta_c\Big|
	= \frac{c}{3}
		\log \sqrt{\frac{c\mspace{1mu}|\mu|}{3a^2}},
}
such that the resulting holographic entanglement entropy reproduces the field theory result with an arbitrary length scale $a$
\eq{
\tilde S[\mA] + \tilde S_W \big[  \tfrac{\ell^2}{a^2}\zeta_c\big]	= S_{a}[\mA].
}

In summary, we have shown that the renormalized entropy on the sphere \eqref{eq:renormalized-entropy} can be obtained from a generalized HRT prescription, where additional endpoint contributions \eqref{eq:entropy-corrections} are included besides the area terms. 
This is the entanglement entropy that is compatible with the partition function \eqref{Zsol} that satisfies the flow equation \eqref{dZdmu}. Our prescription is limited to the \TTbar deformation on the sphere, where we have a concrete result for the replica partition functions for the bulk and the boundary. It would be interesting to consider the endpoint contributions for general backgrounds, which we leave for future study.


\section{General formula for a small interval on the cylinder} \label{se:general-cylinder}

In section \ref{s5} we considered the glue-on HRT proposal for spacetimes with a compact spatial coordinate but zero angular momentum. The methods developed in the main text can be readily applied to more general spacetimes with angular momentum, and we shall summarize the main results in this appendix.

According to the glue-on AdS$_3$ proposal described in section \ref{se:glueonreview}, a \TTbar-deformed CFT with left and right-moving temperatures $T_{L,R}$ can be thought of as living on a cutoff surface in a glue-on version of the rotating BTZ black hole whose metric is given by 
\eq{
ds^2 = \ell^2 \bigg( \frac{d\rho^2}{4 \rho^2} + \frac{ \big( du + \rho \, T_v^2 \, dv \big) \big( dv + \rho \, T_u^2 \, du \big) }{\rho} \bigg) , \qquad \rho \ge \rho_c.
\label{banados}
}
where $\rho$ takes negative values in the $\mathrm{BTZ}^*$ region. The parameters $T_{u,v}$ are given in terms of the inverse temperatures $\beta_{u,v}$ of the BTZ black hole by $T_{u,v} = \pi / \beta_{u,v}$. In analogy with the nonrotating case, they are related to the temperatures of the \TTbar-deformed CFT by \cite{Apolo:2023vnm}
\eq{
	T_u + T_v = \pi\ell\,\frac{T_L + T_R}{\sqrt{1 - (4\pi^2 c\mu / 3)\,T_L T_R}},
\qquad
	T_u - T_v = \pi\ell\,\big(T_L - T_R\big).
\label{Tbulk2}
}
The cutoff surface $\rho = \rho_c$ in these coordinates is related to the deformation parameter by
\eq{
    \ell^{-2} g_{\varphi\varphi}\big|_{\rho = \rho_c}
    \equiv \zeta_c^{-1}
    = -\frac{3\ell^2}{c\mu}.
}

Let us now consider an interval $\mA$ whose endpoints are parametrized in terms of the lightcone coordinates $(u,v)$ by
\eq{
\p\mA = \bigg\{
	\biggl(-\frac{\ell_u}{2} , -\frac{\ell_v}{2} \biggr),
	\bigg(\,\frac{\ell_u}{2} ,  \frac{\ell_v}{2} \bigg)
\bigg\}.
\label{endpointsThermal}
}
The extremal surface $X_\mA$ anchored at these endpoints can be obtained from analytic continuation of a standard HRT surface $X$ anchored at the endpoints of the following auxiliary interval $\hat{\mA}$ at the asymptotic boundary
\eq{
\p\hat{\mA} = \bigg\{
	\biggl(-\frac{\hat{\ell}_u}{2} , -\frac{\hat{\ell}_v}{2} \biggr),
	\bigg(\,\frac{\hat{\ell}_u}{2} ,  \frac{\hat{\ell}_v}{2} \bigg)
\bigg\}.
\label{auxThermal}
}
The candidate HRT surface is thus given by
\eq{
X_{\mA}\colon
\left\lbrace
\begin{aligned}
\ \,
    &\frac{\sinh 2T_u u}{\sinh T_u \hat{\ell}_u}
    = \frac{\sinh 2T_v v}{\sinh T_v \hat{\ell}_v},
\\[1ex]
    &\rho = \frac{1}{T_u T_v}\,
        \frac{\sinh T_v \hat{\ell}_v}{\sinh T_u \hat{\ell}_u}\,
        \frac{\cosh T_u \hat{\ell}_u - \cosh 2T_u u}
            {\cosh T_v \hat{\ell}_v + \cosh 2T_v v}\,,
\end{aligned}
\right.\qquad \rho \ge \rho_c.
\label{eq:RTcutoffTemp}
}
where the auxiliary lengths $\hat \ell_{u,v}$ can be related to the physical ones $\ell_{u,v}$ via
\eq{
\begin{aligned}
    \frac{\cosh T_u \hat{\ell}_u}{\cosh T_u \ell_u}
    &= \frac{
            1 - \zeta_c (T_u^2 + T_v^2) + 2\zeta_c T_u T_v
            \,{\tanh T_u \ell_u}
            \big/{\tanh T_v \ell_v}
        }{
            \sqrt{\big( 1 - \zeta_c (T_u^2 + T_v^2) \big)^2 - 4 \zeta_c^2 T_u^2 T_v^2}
        },
\\[.5ex]
    \frac{\cosh T_v \hat{l}_v}{\cosh T_v l_v}
    &= \frac{
            1 - \zeta_c (T_u^2 + T_v^2) + 2\zeta_c T_u T_v
            \,{\tanh T_v \ell_v}
            \big/{\tanh T_u \ell_u}
        }{
            \sqrt{\big( 1 - \zeta_c (T_u^2 + T_v^2) \big)^2 - 4 \zeta_c^2 T_u^2 T_v^2}
        }.
\end{aligned}
\label{eq:infinity_by_cutoff_lengths}
}
In analogy with previous cases, the existence of the gluing points \eqref{auxThermal}, i.e.~the reality of $\hat \ell_{u,v}$, requires the interval to be larger than a minimum value. Assuming that the extremal surface \eqref{eq:RTcutoffTemp} exists, and integrating the signed area along $X_\mA$, we find that its contribution to the glue-on HRT formula is given by
\eq{
    \tilde S_{\ell_u, \ell_v} 
    = \frac{c}{6}\, \biggl(
        \arctanh \bigg[\bigg(
            \frac{\tanh T_u \hat{\ell}_u}{\tanh T_u \ell_u}
        \bigg)^{\!\!\frac{\abs{\mu}}{\mu}}
        \bigg]
        + \arctanh \bigg[\bigg(
            \frac{\tanh T_v \hat{\ell}_v}{\tanh T_v \ell_v}
        \bigg)^{\!\!\frac{\abs{\mu}}{\mu}\,}
        \bigg]
    \biggr),
\label{eq:cylinder-generic-formula}
}
where $T_{u,v}$ and $\hat \ell_{u,v}$ may be written in terms of $T_{L,R}$ and $\ell_{u,v}$ using \eqref{Tbulk2} and \eqref{eq:infinity_by_cutoff_lengths}. 

For an interval $\mA$ whose length is small enough compared to the size of the system, but large enough so that the extremal surface exists, we have $\tilde{S}[\mA] = \tilde S_{\ell_u, \ell_v}$. Let $\ell_u = \ell_v = \ell_\mA$. Then for $T_u = T_v = i/2$ we recover the global $\mathrm{AdS}_3$ result \eqref{HEEglobalAdS}; for $T_u = T_v = \pi/\beta_t$ we recover \eqref{HEEBTZXA} for the nonrotating BTZ black hole; while for $T_u = T_v = 0$ we recover the result for Poincar\'e $\mathrm{AdS}_3$ \eqref{HEEpoincareAdS}, which is locally identical to the massless BTZ black hole. In analogy with the discussion in section \ref{se:ERTBTZ}, it would be interesting to analyze the phase structure in this more general case,  which we leave to future work.


\bibliographystyle{JHEP.bst}
\bibliography{glueon.bib}

\end{document}